\documentclass[12pt, onecolumn, a4paper]{article}
\usepackage{amsmath,amssymb,cite,color,epsfig,float,footmisc,geometry,graphicx,mathtools,multirow,mwe,setspace,soul,svg,xcolor}
\usepackage[utf8]{inputenc}
\usepackage{tcolorbox}
\begin{document}

\thispagestyle{empty}

\begin{center}

$ $

\bigskip
\bigskip

\LARGE \textbf{White Paper on Broadband Connectivity in 6G}

\bigskip
\bigskip

\large Nandana~Rajatheva,\footnote{\label{cwc}Centre for Wireless Communications, University of Oulu, Finland (email: \{nandana.rajatheva, italo.atzeni, marcos.katz, mohammadjavad.salehi, nuwanthika.rajapaksha,  antti.tolli\}@oulu.fi).} Italo~Atzeni,\footref{cwc} Emil~Björnson,\footnote{Department of Electrical Engineering (ISY), Linköping University, Sweden (email: emil.bjornson@liu.se).} André~Bourdoux,\footnote{IMEC, Belgium (email: andre.bourdoux@imec.be).} Stefano~Buzzi,\footnote{Department of Electrical and Information Engineering, University of Cassino and Southern Latium, Italy (email: buzzi@unicas.it).} Jean-Baptiste~Doré,\footnote{\label{cea}CEA-Leti, France (email: jean-baptiste.dore@cea.fr).} Serhat~Erkucuk,\footnote{\label{khu}Department of Electrical and Electronics Engineering, Kadir Has University, Turkey (email:\{serkucuk, eepanay\}@khas.edu.tr).} Manuel~Fuentes,\footnote{Institute of Telecommunications and Multimedia Applications, Universitat Politecnica de Valencia, Spain (email: mafuemue@iteam.upv.es).} Ke~Guan,\footnote{\label{bjtu}State Key Laboratory of Rail Traffic Control and Safety, Beijing Jiaotong University, and Beijing Engineering Research Center of High-speed Railway Broadband Mobile Communications, China (email:kguan@bjtu.edu.cn).} Yuzhou~Hu,\footnote{Algorithm Department, ZTE Corporation, China (email: hu.yuzhou@zte.com.cn).} Xiaojing~Huang,\footnote{School of Electrical and Data Engineering, Faculty of Engineering and Information Technology, University of Technology Sydney, Australia (email: xiaojing.huang@uts.edu.au).} Jari~Hulkkonen,\footnote{\label{nokia}Nokia Bell Labs, Finland (email: \{jari.hulkkonen, oskari.tervo\}@nokia-bell-labs.com).} Josep~Miquel~Jornet,\footnote{Department of Electrical and Computer Engineering, Institute for the Wireless Internet of Things, Northeastern University, USA (email: j.jornet@northeastern.edu).} Marcos Katz,\footref{cwc} Rickard~Nilsson,\footnote{Department of Computer Science, Electrical and Space Engineering, Luleå University of Technology, Sweden (email: rickard.o.nilsson@ltu.se).} Erdal~Panayirci,\footref{khu} Khaled~Rabie,\footnote{Department of Engineering, Manchester Metropolitan University, UK (email: k.rabie@mmu.ac.uk).} Nuwanthika~Rajapaksha,\footref{cwc} MohammadJavad~Salehi,\footref{cwc} Hadi~Sarieddeen,\footnote{Division of Computer, Electrical and Mathematical Sciences and Engineering, King Abdullah University of Science and Technology, Saudi Arabia (email: hadi.sarieddeen@kaust.edu.sa).} Tommy~Svensson,\footnote{Department of Electrical Engineering, Chalmers University of Technology, Sweden (email: tommy.svensson@chalmers.se).} Oskari~Tervo,\footref{nokia} Antti~Tölli,\footref{cwc} Qingqing~Wu,\footnote{Department of Electrical and Computer Engineering, National University of Singapore, Singapore (email: wu.qq1010@gmail.com).} and~Wen~Xu\footnote{Huawei Technologies, Germany (email: wen.dr.xu@huawei.com).}

\makeatletter{\renewcommand*{\@makefnmark}{}
\footnotetext{This paper is a compilation of ideas presented by various entities at the 6G Wireless Summit 2020. It does not reflect an agreement on all the included technology aspects by all the involved entities.}\makeatother}

\bigskip
\bigskip

\large \today
\end{center}

\newpage

%=========================================================================
\section*{Executive Summary}
%=========================================================================

This white paper explores the road to implementing broadband connectivity in future 6G wireless systems. Different categories of use cases are considered, from extreme capacity with peak data rates up to 1~Tbps, to raising the typical data rates by orders-of-magnitude, to support broadband connectivity at railway speeds up to 1000 km/h. To achieve these goals, not only the terrestrial networks will be evolved but they will also be integrated with satellite networks, all facilitating autonomous systems and various interconnected structures.

We believe that several categories of enablers at the infrastructure, spectrum, and protocol/algorithmic levels are required to realize the intended broadband connectivity goals in 6G. At the infrastructure level, we consider ultra-massive MIMO technology (possibly implemented using holographic radio), intelligent reflecting surfaces, user-centric and scalable cell-free networking, integrated access and backhaul, and integrated space and terrestrial networks. At the spectrum level, the network must seamlessly utilize sub-6 GHz bands for coverage and spatial multiplexing of many devices, while higher bands will be used for pushing the peak rates of point-to-point links. The latter path will lead to THz communications complemented by visible light communications in specific scenarios. At the protocol/algorithmic level, the enablers include improved coding, modulation, and waveforms to achieve lower latencies, higher reliability, and reduced complexity. Different options will be needed to optimally support different use cases. The resource efficiency can be further improved by using various combinations of full-duplex radios, interference management based on rate-splitting, machine-learning-based optimization, coded caching, and broadcasting. Finally, the three levels of enablers must be utilized not only to deliver better broadband services in urban areas, but also to provide full-coverage broadband connectivity must be one of the key outcomes of 6G.

\newpage

\tableofcontents

\newpage

\onehalfspacing

%=========================================================================
\section{Introduction}
%=========================================================================

\definecolor{apricot}{rgb}{0.98, 0.81, 0.69}
\begin{tcolorbox}[width=\textwidth,colback=apricot,outer arc=0mm]    
  {\bf  What will beyond-5G and 6G wireless networks look like? Which new technology components will be at their heart? Will the 1~Tbps frontier be reached?
}
\end{tcolorbox}    

Even though these questions are hard to answer with certainty at this time, the research community has started to look for the answers and some interesting concepts have already emerged. A conservative and cautious answer is that \textit{6G networks will be based on technologies that were not yet mature for being included in 5G systems}. Based on this argument, every technology that will not be in 3GPP standards by the end of 2020 will be a possible ingredient of future generations of wireless cellular systems. 
A more daring answer is that \textit{6G networks will be based on new technologies that were not-at-all considered when designing and developing 5G networks, combined with enhancements of technologies that were already present in the previous generation of wireless cellular networks}. Based on the latter statement, we can certainly state that 6G wireless systems will:
\begin{itemize}
\item[(i)] Be based on \textbf{extreme densification of the network infrastructure}, such as access points (APs), which will cooperate to form a cell-free network with seamless quality of service over the coverage area;

\item[(ii)] Make intense use of \textbf{distributed processing and cache memories}, for example, in the form of the cloud-RAN technology;

\item[(iii)] Continue the trend of complementing the wide-area coverage achieved at sub-6 GHz frequencies with \textbf{using higher and higher carrier frequencies}, beyond millimeter wave (mmWave) and up to the visible light band, to provide high-capacity point-to-point links;

\item[(iv)] Leverage \textbf{network slicing and multi-access edge computing} to enable the birth of new services with specialized performance requirements and to provide the needed resources to support vertical markets;

\item[(v)] Witness an increasing \textbf{integration of terrestrial and satellite wireless networks}, with a big role played by unmanned-aerial-vehicles (UAVs) and low-earth-orbit (LEO) micro-satellites, to fill coverage holes and offload the network in heavy-load situations; 

\item[(vi)] Leverage \textbf{machine learning methodologies} to improve the complexity and efficiency of traditional model-based algorithms for signal processing and resource allocation. \end{itemize}

This white paper has been written by researchers from academia, research centers, industry, and standardization bodies. It extends the work in \cite{6GFlagship_WP} with a more specific focus on broadband connectivity, which might be the most important use case in 6G (though far from the only one). In particular, it describes the physical (PHY) layer and medium-access control (MAC) layer methodologies for achieving 6G broadband connectivity with very high data rates, up to the Tbps range.

We begin by briefly presenting different use cases for the broadband technology, key performance indicators, and spectrum ranges for 6G. Then, we survey all the expected enablers for achieving 6G broadband access. These include:
\begin{itemize}
\item \textbf{Enablers at the infrastructure level} based on the evolution of massive multiple-input multiple-output (MIMO) and holographic radio, intelligent reflecting surfaces, user-centric and scalable cell-free networking, integrated access and backhaul, and integrated space and terrestrial networks;

\item \textbf{Enablers at the spectrum level} to realize THz communications and visible light communications (VLCs);

\item \textbf{Enablers at the protocol/algorithmic level}, providing innovations in the field of coding, modulation, waveform, and duplex, interference management using non-orthogonal multiple access (NOMA) and rate splitting, machine learning-aided algorithms, coded caching, wideband broadcast, and full-coverage broadband connectivity.
\end{itemize}
Finally, we conclude with an overview of the main research and technological challenges that must be overcome to reach our vision of 6G wireless systems.

%=========================================================================
\section{Use Cases, Key Performance Indicators, and Spectrum} \label{sec:use_cases}
%=========================================================================

%-------------------------------------------------------------------------
\subsection{Use Cases}

\begin{figure}[t]
\centering
\includegraphics[scale=0.75]{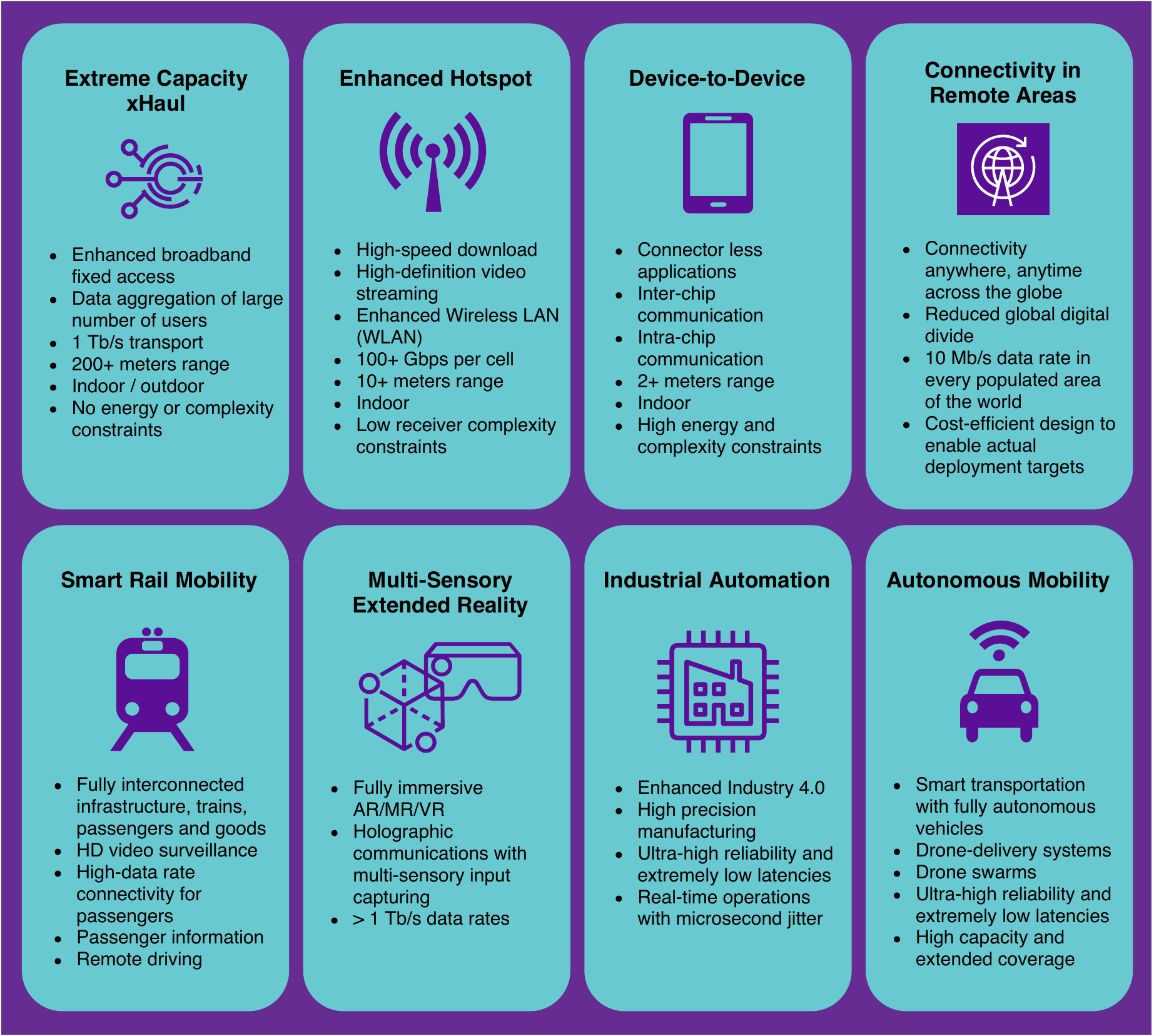}
\caption{Different 6G use cases and applications.}
\label{fig:Use1}
\end{figure}

The emergence and need for 6G technology will be governed by the unprecedented performance requirements arising from exciting new applications foreseen in the 2030 era, which existing cellular generations will not be able to support. This white paper focuses on applications that require broadband connectivity with high data rates, in combination with other specialized characteristics. The following is a list of the potential new use cases and applications in 6G which will help to understand the key requirements of future 6G systems. A summary of those discussed use cases is presented in Figure \ref{fig:Use1}.

\begin{itemize}
\item \textbf{Extreme capacity xHaul:} This use case refers to a fixed symmetric point-to-point link targeting high data rates without energy or complexity constraints since no user devices are involved. This can only be enabled using a combination of high bandwidth and high spectral efficiency. The envisioned ultra-dense network topology in urban areas with extreme capacity and latency requirements makes fiber-based backhauling highly desirable but complicated due to the limited fiber network penetration (variable from country to country) and related extension cost. Hence, wireless infrastructure is needed as a flexible complement to optical fiber deployment, both indoor and outdoor, to avoid any bottleneck in the backhaul (or xHaul). Ultra-high speed is required since backhaul aggregates the data rates of many user devices. The xHaul can also provide efficient access to computing resources at the edge or in the cloud. Figure \ref{fig:Use2} depicts an example setup of an extreme capacity xHaul network in an indoor environment.

\begin{figure}[t]
\centering
\includegraphics[scale=0.6]{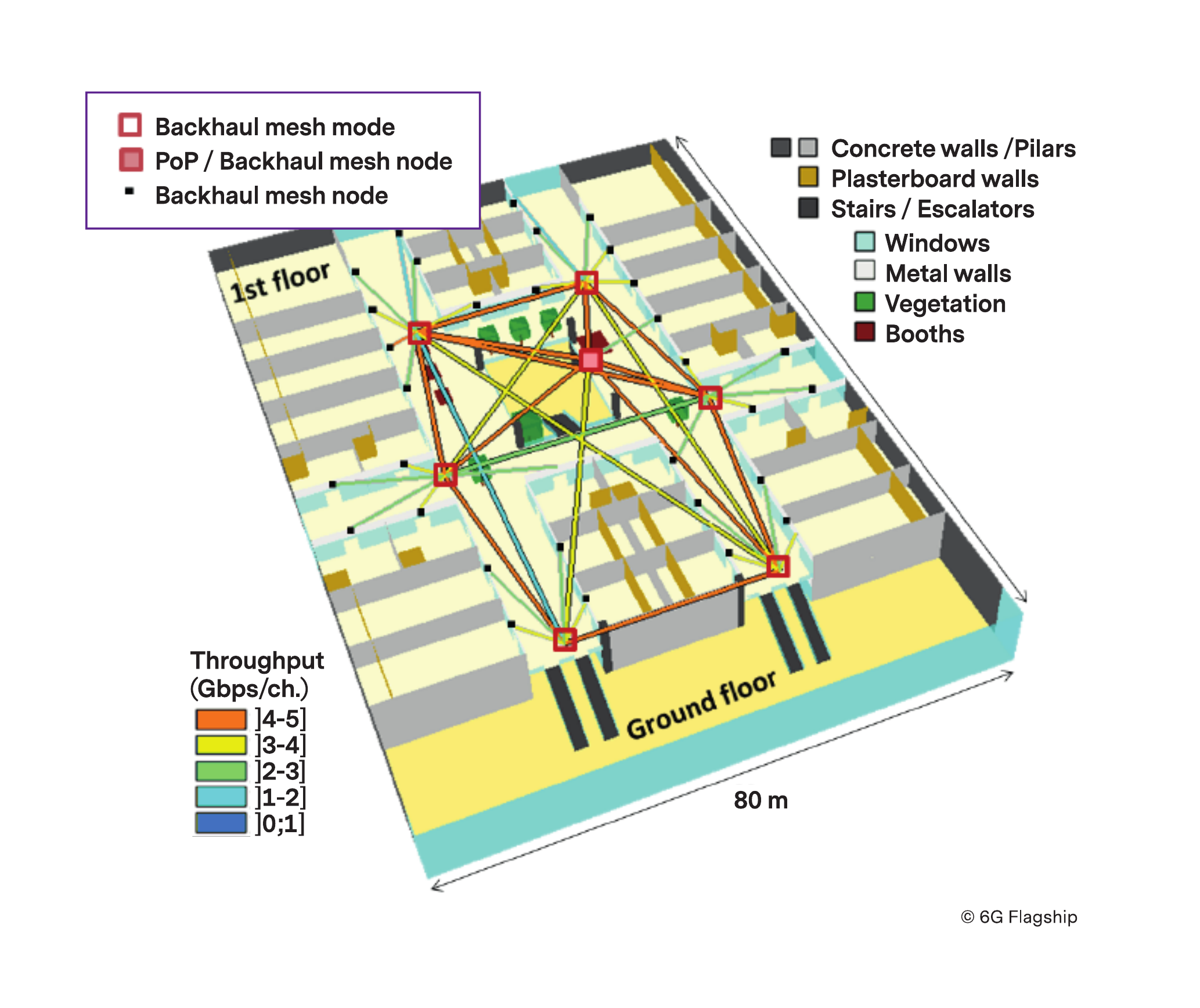}

\vspace{-8mm}

\caption{An example setup of an extreme capacity xHaul network in an indoor environment \cite{EucapSiradelCea}.}
\label{fig:Use2}
\end{figure}

\item \textbf{Enhanced hotspot:} An enhanced hotspot entails a high-rate downlink from the AP to several user devices, with short coverage and low receiver complexity constraints. Envisaged applications are hotspots delivering high-speed data to demanding applications such as high-definition video streaming and enhanced Wireless LAN (WLAN).

\item \textbf{Short-range device-to-device communications:} A symmetric high-rate point-to-point link with very stringent energy and complexity constraints is considered here. The use case focuses on data exchange between user devices that are within a short distance, with limited involvement of the network infrastructure. This use case also includes inter/intra-chip communications and wireless connectors, among others.

\item \textbf{Smart rail mobility:} This is a required component in a paradigm where infrastructure, trains, passengers, and goods are seamlessly connected with high data rates. Railway communications are evolving from only supporting critical signaling applications to providing several bandwidth-intensive applications, such as on-board and wayside high-definition video surveillance, broadband connectivity for passengers, broadcasting of passenger information, and remote driving or control. These applications need to be deployed in at least five scenarios: train-to-infrastructure, inter-wagon (the train backbone), intra-wagon, inside the station, and infrastructure-to-infrastructure.

\item \textbf{Multi-sensory extended reality:} Augmented, mixed, and virtual reality (AR/MR/VR) applications, capturing multi-sensory inputs and providing real-time user interaction are considered under this use case. Very high per-user data rates in Gbps range and very low latencies are required to deliver a fully immersive experience \cite{9040264_Marco}. Remote connectivity and interaction powered by holographic communications, along with all human sensory input information, will further push the data rate and latency targets. Multiple-view cameras used for holographic communications will require data rates on the order of terabits per second \cite{8792135_Emilio}.

\item \textbf{Industrial automation and robotics:} Industry 4.0 envisions a digital transformation of manufacturing industries and processes through cyber-physical systems, internet-of-things (IoT) networks, cloud computing, and artificial intelligence. In order to achieve high-precision manufacturing, automatic control systems, and communication technologies are utilized in the industrial processes. Ultra-high reliability on the order of $1$-$10^{-9}$ and extremely low latency around 0.1-1 ms round-trip time are expected in communications, along with real-time data transfer with guaranteed microsecond delay jitter in industrial control networks \cite{8792135_Emilio}. While 5G initiated the implementation of Industry 4.0, 6G is expected to uncover its full potential supporting those stringent requirements by the novel, disruptive technologies brought by 6G.

\item \textbf{Autonomous mobility:} The smart transportation technologies initiated in 5G are envisioned to be further improved towards fully autonomous systems, providing safer and efficient transportation, efficient traffic management, and improved user experience. Connected autonomous vehicles demand reliability above 99.99999\% and latency below 1 ms, even in very high mobility scenarios up to 1000 km/h \cite{8792135_Emilio, 9040264_Marco}. Moreover, higher data rates are required due to the increased number of sensors in vehicles that are needed to assist autonomous driving. Other autonomous mobility solutions like drone-delivery systems and drone swarms are also evolving in different application areas such as construction, emergency response, military, etc. and require improved capacity and coverage \cite{9040264_Marco}.

\item \textbf{Connectivity in remote areas:} Half of the world's population still lacks basic access to broadband connectivity. The combination of current technologies and business models have failed to reach large parts of the world. To reduce this digital divide, a key target of 6G is to guarantee 10 Mbps in every populated area of the world, using a combination of ground-based and spaceborne network components. Importantly, this should not only be theoretically supported by the technology but 6G must be designed in a sufficiently cost-efficient manner to enable actual deployments that deliver broadband to the entire population of the world.

\item \textbf{Other use cases:} Some other applications that 6G is expected to enable or vastly enhance include: internet access on planes, wireless brain-computer interface based applications,
broadband wireless connectivity inside data centers, Internet of Nano-Things and Internet of Bodies through smart wearable devices and intrabody communications achieved by implantable nanodevices and nanosensors \cite{8766143_Zhengquan}.

\end{itemize}

%-------------------------------------------------------------------------
\subsection{Key Performance Indicators}

The key performance indicators (KPIs) have to a large extent stayed the same in several network generations \cite{ITU-IMTA,ITU-IMT2020}, while the minimum requirements have become orders-of-magnitude sharper. One exception is the energy efficiency, which was first introduced as a KPI in 5G, but without specifying concrete targets. We believe 6G will mainly contain the same KPIs as previous generations but with greatly higher ambitions. However, while the KPIs were mostly independent in 5G (but less stringent at high mobility and over large coverage areas), a cross-relationship is envisaged in 6G through a definition of groups. All the indicators in a group should be fulfilled at the same time, but different groups can have diverse requirements. The reason is that we will move from a situation where broadband connectivity is delivered in a single way to a situation where the requirements of different broadband applications become so specialized that the union of them cannot be simultaneously achieved. Hence, 6G will need to be real-time configurable to cater to these different groups.

The following are the envisaged KPIs.

\begin{itemize}
\item \textit{Extreme data rates}: Peak data rates up to 1~Tbps are envisaged for both indoor and outdoor connectivity. The user-experienced data rate, which is guaranteed to 95\% of the user locations, is envisioned to reach 1 Gbps.

\item \textit{Enhanced spectral efficiency and coverage}: The peak spectral efficiency can be increased using improved MIMO technology and modulation schemes, likely up to 60 b/s/Hz. However, the largest envisaged improvements are in terms of the uniformity of the spectral efficiency over the coverage area. The user-experienced spectral efficiency is envisaged to reach 3 b/s/Hz. Moreover, new PHY layer techniques are needed to allow for broadband connectivity in high mobility scenarios and more broadly the scenarios for which former wireless networks generations do not fully meet the needs. 

\item \textit{Very wide bandwidths}: To support extremely high peak rates, the maximum supported bandwidth must greatly increase. Bandwidths up to 10 GHz can be supported in mmWave bands, while up to 100 GHz can be reached in THz and visible light bands.

\item \textit{Enhanced energy efficiency}: Focusing on sustainable development, 6G technologies are expected to pay special attention in achieving better energy efficiency, both in terms of the absolute power consumption per device and the transmission efficiency. In the latter case, the efficiency should reach up to  1 terabit per Joule. Hence, developing energy-efficient communication strategies is a core component of 6G.

\item \textit{Ultra-low latency}: The use of bandwidths that are wider than 10 GHz will allow for latency down to 0.1 ms. The latency variations (jitter) should reach down to 1 $\mu$s, to provide an extreme level of determinism.

\item \textit{Extremely high reliability}: Some new use cases require extremely high reliability up to $1$-$10^{-9}$ to enable mission and safety-critical applications. 

\end{itemize} 

\begin{figure}[t]
\includegraphics[scale=0.75]{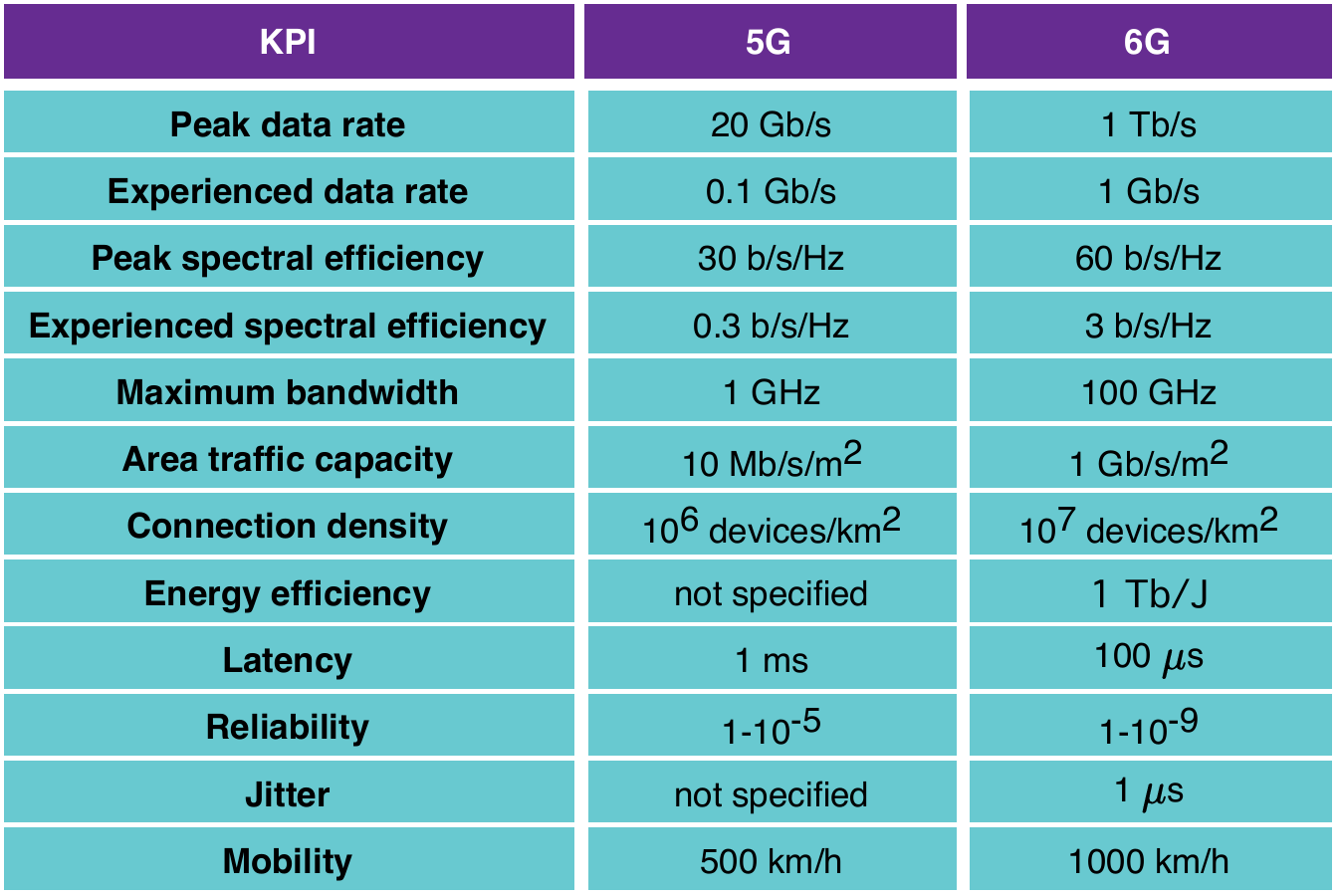}
\centering
\caption{A comparison of 5G and 6G KPIs \cite{8766143_Zhengquan, 8792135_Emilio, 9040264_Marco,Bjornson2018b}.}
\label{fig:kpi}
\end{figure}

It is unlikely that all of these requirements will be simultaneously supported, but different use cases will have different sets of KPIs, whereof only some reach the maximum requirements mentioned above. A comparison of the 5G and 6G KPIs is shown in Figure \ref{fig:kpi}, where also the area traffic capacity and connection density are considered.

It is very likely that 6G will to a large extent carry information related also to non-traditional applications of wireless communications, such as distributed caching, computing, and AI decisions. Thus, we need to investigate whether there is a need to introduce new KPIs for such applications, or if the traditional KPIs are sufficient.

%-------------------------------------------------------------------------
\subsection{6G Spectrum}

5G defines separately operations for sub-6 GHz and 24.25 to 52.6 GHz. 
The band from 57 to 71 GHz is already in 3GPP Rel-17 agenda extending the upper limit to 71 GHz \cite{RP-193259} and band options up to 114.25 GHz were included in Release 16 pre-study on New Radio beyond 52.6 GHz. 

\begin{figure}[t]
\centering
\includegraphics[width=14cm]{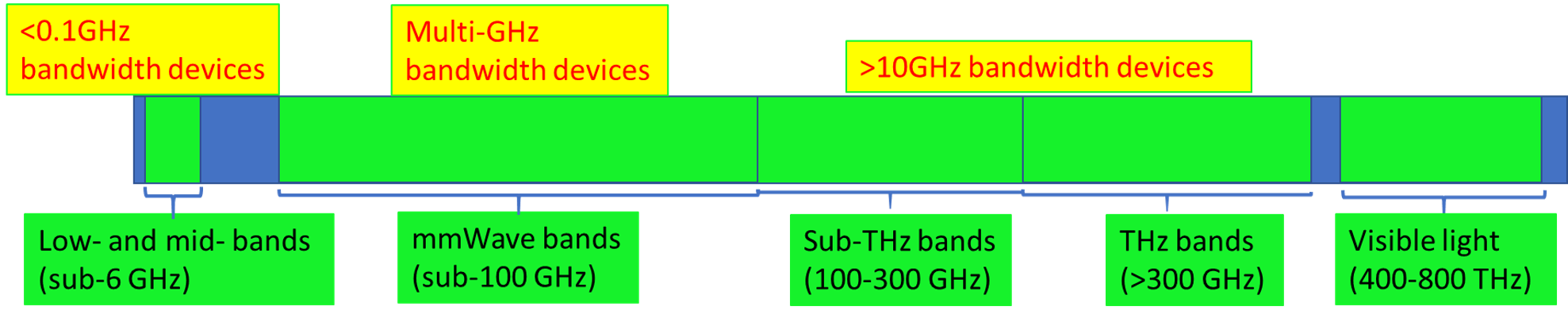}
\caption{Potential spectrum regions for 6G.}
\label{fig:SpectrumBands}
\end{figure}

In the 6G era, we expect to see enhancements over the current 5G spectrum, and expansion of the spectrum to potential new bands from low-bands to low THz and visible light region. The potential spectrum regions are illustrated in Figure \ref{fig:SpectrumBands}. The potential communication bandwidth is expected to increase when going to higher frequencies.
For example, up to $18$ GHz aggregated bandwidth is available for fixed communications in Europe in the frequency band $71-100$ GHz, while in the USA, both mobile and fixed communications are allowed. There are also bands between $95$ GHz and $3$ THz recently opened by the Federal Communications Commission (FCC) for experimental use to support the development of higher bands \cite{6G2020Tervo}.
When going to higher frequencies, the intention is not to achieve a gradual increase in operational frequency but a convergence of existing technologies in these different bands into a joint wireless interface that enables seamless handover between bands.

The operation in existing bands will be enhanced in 6G with respect to the KPIs described above, but not all targets are expected to be reached in all frequency bands. For example, low frequency bands are often preferable in terms of spectral efficiency, reliability, mobility support, and connectivity density. In contrast, high frequency bands are often preferable in terms of peak data rates and latency. 

There is a long experience in operating wireless communication systems in the sub-6 GHz spectrum.
Moving from sub-6 GHz to mmWave introduced several technical challenges from initial access to beamforming implementation since fully digital solutions take time to develop. The development of 5G has led to large innovations in these respects. It becomes even more challenging when going to higher frequencies. The main reason for moving to higher frequency bands is to greatly expand the available spectrum. In the sub-THz to THz region, hardware constraints such as the speed of data converters and computational complexity will cause challenges to make efficient use of wide bandwidths. Moreover, investigations into new waveforms, mitigation of hardware impairments, as well as new materials to implement devices operating in that spectrum are required. This makes the sub-THz to THz region a new area where many open research problems exist from the hardware to the physical layer protocols.

Last but not least, the deployment of cellular technologies in unlicensed bands will continue and could be a game-changer in the next generation of wireless systems.  

%=========================================================================
\section{Enablers at the Infrastructure Level}
%=========================================================================

%-------------------------------------------------------------------------
\subsection{Ultra-Massive MIMO and Holographic Radio}

Massive MIMO is a cellular technology where the access points are equipped with a large number of antennas, which are utilized for spatial multiplexing of many data streams per cell to one or, preferably, multiple users.
The massive MIMO technology has become synonymous with 5G, but the hardware implementation and algorithms that are used differ to a large extent from what was originally proposed in \cite{Marzetta2010a} and then described in textbooks on the topic \cite{Marzetta2016a,massivemimobook}.
For example, compact 64-antenna rectangular panels with limited angular resolution in the azimuth and elevation domains are being used instead of physically large horizontal uniform linear arrays with hundreds of antennas, which would lead to very narrow azimuth beams.
Moreover, a beam-space approach is taken where 2D discrete Fourier transform (DFT) codebooks are used to describe grids of 64 beams in predetermined angular directions for rectangular panels, while only one of these 64 predefined beams is selected for each user. This approach is only appropriate for line-of-sight (LOS) communication with calibrated planar arrays and widely spaced users. In general, non-LOS channels contain arbitrary linear combinations of those beams, the arrays might have different geometry, and the array responses of imperfectly calibrated arrays are not described by a DFT. A practical reason for these design simplifications is that analog and hybrid beamforming were needed to come quickly to the market in 5G. However, fully digital arrays will be available for a wide range of frequencies (including mmWave) when 6G arrives and, therefore, we should utilize it to implement something capable of providing a performance substantially closer to what the massive MIMO theory suggests \cite{Marzetta2016a,massivemimobook}.
Since the massive MIMO terminology has become diluted by many suboptimal design choices in 5G, we will use the term \emph{ultra-massive MIMO}~\cite{akyildiz2016realizing} in this paper to describe the 6G version of the technology.

\subsubsection{Beamforming beyond the Beam-Space Paradigm}

The spectrum resources are scarce, particularly in the sub-6 GHz bands that will always define the baseline coverage of a network. Hence, achieving full utilization of the spatial dimensions (i.e., how to divide the available spatial resources between concurrent transmissions) is very important. 
The beamforming from antenna arrays has traditionally been based on a beam-space model, where the spatial domain is described by a small set of predefined beams. Each beam represents a transmission in a particular angular direction. Current planar arrays are capable of generating a set of beams with varying azimuth and elevation angles. This is usually called full-dimensional beamforming, although it is far from utilizing all available spatial dimensions.
This traditional beam-space paradigm describes the signal propagation in three dimensions, while a 64-antenna array is capable of creating beams in a 64-dimensional vector space where most beams lack a clear angular directivity but utilize the multipath environment to focus the signal at certain points in space. It is only in propagation with no near-field scattering, a pre-defined array geometry, perfectly calibrated arrays, and no mutual coupling, where the low-dimensional approximation provided by the 2D-DFT-based beam-space approach can be made without loss of optimality.
These restrictions are impractical when considering ultra-massive MIMO arrays that are likely to interact with devices and scattering objects in the near-field, can have arbitrary geometry, and will likely also feature imperfect hardware calibration and coupling effects. That is why the massive MIMO concept was originally designed to use uplink pilots to estimate the entire channel instead of its three-dimensional far-field approximation \cite{Marzetta2016a,massivemimobook}.

An alternative way to increase the dimension is to start from a three-dimensional approximation and make use of orbital angular momentum (OAM) to raise the dimension. This method makes use of the ``vortex mode'' to distinguish between users having approximately the same angular direction, but this is also just a lower-dimensional approximation of the true channel \cite{OAMedfors}. In other words, there is no need to explicitly make use of OAM in 6G since the original massive MIMO is already implicitly doing it.

The beamforming challenge for 6G is to make use of physically large panels, since the dimensionality of the beamforming is equal to the number of antennas, and the beamwidth is inversely proportional to the array aperture. With an ultra-high spatial resolution, each transmitted signal can be focused in a small region around the receiver, leading to a beamforming gain proportional to the number of antennas. An ultra-high level of spatial multiplexing is also possible since each signal is only strong in a small region around the receiver. This is particularly important to make efficient use of the low bands, where the channel coherence time is large and, therefore, can accommodate the channel estimation overhead for many users. With a sufficient number of antennas, 1 MHz of spectrum in the 1 GHz band can give the same data rate as 100 MHz of spectrum in the 100 GHz band. The reason is that the lower band supports spatial multiplexing of 100 times more users since the coherence time is 100 times larger.

Ideally, ultra-massive MIMO should be implemented using fully digital arrays with hundreds or thousands of phase-synchronized antennas. 
This is practically possible in both sub-6 GHz and mmWave bands \cite{Bjornson2019d}, but the implementation complexity grows with the carrier frequency. On the one hand, new implementation concepts are needed that are not stuck in the suboptimal beam-space paradigm, as in the case of hybrid beamforming, but can make use of all the spatial dimensions. On the other hand, new device technologies, potentially leveraging new materials, can be utilized to implement on-chip compact ultra-massive antenna arrays that can potentially enable fully digital architectures~\cite{singh2020operation_tx}. Doubly massive MIMO links, wherein a large number of antennas is present at both sides of the communication link, will also be very common at mmWave frequencies \cite{Buzzi_DMIMO2,8269171}.

Continuous aperture antennas are considered for improving beamforming accuracy. While the beamwidth of the main-lobe is determined by the array size, the continuous aperture gives cleaner transmissions with smaller side-lobes \cite{Asilomar2019a}. To implement a continuous aperture antenna array, one method is to use a dense array of conventional discretely spaced antennas, but the cost and energy consumption will be high if every element should have an individual radio-frequency (RF) chain. Another solution is to integrate a large number of antenna elements into a compact space in the form of a meta-surface. However, this method is limited to the passive setup described in the next section, because for a continuous-aperture active antenna array, the RF feed network is simply impossible to achieve due to the ultra-dense elements. A possible solution is the holographic radio technology.

\subsubsection{Holographic Radio}

Holographic radio is a new method to create a spatially continuous electromagnetic aperture to enable holographic imaging-level, ultra-high density, and pixelated ultra-high resolution spatial multiplexing \cite{Zong20196G}. 
In general, holography records the electromagnetic field in space based on the interference principle of electromagnetic waves. The target electromagnetic field is reconstructed by the information recorded by the interference of reference and signal waves. The core of holography is that the reference wave must be strictly coherent as a reference, and the holographic recording sensor must be able to record the continuous wave-front phase of the signal wave so as to record the high-resolution holographic electromagnetic field accurately. Because radio and light waves are both electromagnetic waves, holographic radios are very similar to optical holography \cite{Bjornson2019d}. For holographic radios, the usual holographic recording sensor is the antenna.

\vspace{\baselineskip} 
\noindent \textbf{Realization of holographic radio}

To achieve a continuous aperture active antenna array, an ingenious method is to use an ultra-broadband tightly coupled antenna array (TCA) based on a current sheet. The uni-traveling-carrier photodetectors (UTC-PD) are bonded to antenna elements through flip-chip technology and form the coupling between antenna elements \cite{Konkol2017High}. In addition, the patch elements are directly integrated into the electro-optic modulator. The current output by UTC-PD directly drives the antenna elements, so the entire active antenna array has a very large bandwidth (about 40~GHz). Moreover, this innovative continuous-aperture active antenna array does not require an ultra-dense RF feed network at all, which not only has achievability but also has clear implementation advantages.

\begin{figure}[t]
\includegraphics[scale=0.7]{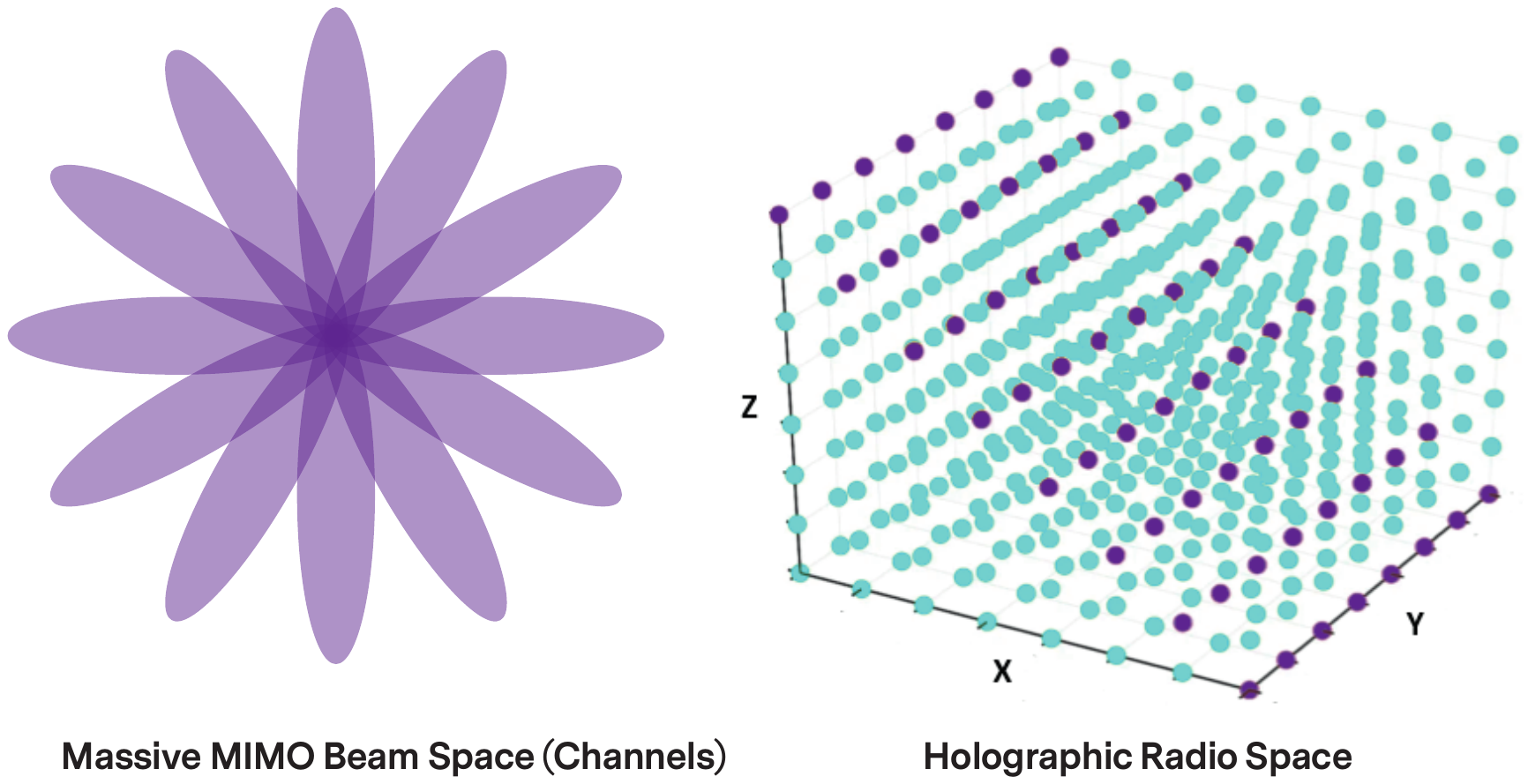}
\centering \label{fig:radioSpace}
\caption{Comparison between beam space (channel) in massive MIMO and radio space in holographic radio}
\end{figure}

Unlike the beam-space approach to massive MIMO, which has dominated in 5G, holographic radios are capable of making use of many more spatial dimensions by making use of a diffraction model based on Huygens' principle.
The signal wavefront is a near-plane wave, so there is no concept of beams, only interference patterns, that is, a holographic radio space. Correspondingly, accurate computation of communication performance requires detailed electromagnetic numerical computations for radio space, that is, algorithms and tools related to computational electromagnetics and computational holography. The spatial channel correlation is described based on the Fresnel-Kirchoff integral. Moreover, holographic radio uses holographic interference imaging to obtain the RF spectral hologram of the RF transmitting sources (UEs), not requiring CSI information or channel estimation. At the same time, a 3D constellation of distributed UEs in the RF phase space can be obtained through spatial spectral holography, providing precise feedback for spatial RF wave field synthesis and modulation in the downlink. Spatial RF wave field synthesis and modulation can obtain a 3D pixel-level structured electromagnetic field (similar to an amorphous or periodic electromagnetic field lattice), which is the high-density multiplexing space of holographic radio and different from sparse beam space considered in 5G.

It is noteworthy that all waveforms that can be generated using a holographic array can also be generated using an appropriately shaped array of half-wavelength-spaced discrete antennas. Hence, the classical theory for channel estimation and signal processing in massive MIMO \cite{massivemimobook} can, in principle, be used to achieve the same performance. However, since the holographic radio is utilizing the advantages of optical processing, spectrum computing, large-scale photon integration, electro-optical mixing, and analog-digital photon hybrid integration technologies, a new physical layer must be designed to make use of these new methods.

\vspace{\baselineskip} 
\noindent \textbf{Signal processing for holographic radio}

There are different ways to implement holographic radios for the purpose of joint imaging, positioning, and wireless communications \cite{Xu2017Holographic}. However, extreme broadband spectrum and holographic RF generation and sensing will generate massive amounts of data. These are challenging to process to perform critical tasks with low-latency and high reliability. Thus, machine learning might be required to operate the system effectively. To meet the 6G challenges of energy efficiency, latency, and flexibility, a hierarchical heterogeneous optoelectronic computing and signal processing architecture will be an inevitable choice \cite{Baiqing2017Photonics}. Fortunately, holographic radios achieve ultra-high coherence and high parallelism of signals by coherent optical up-conversion of the microwave photonic antenna array, and this ultra-high coherence and high parallelism also facilitate the signal to be processed directly in the optical domain. However, it is challenging to adapt the signal processing algorithms of the physical layer to fit the optical domain.

How to realize holographic radio systems is a wide-open area. Due to the lack of existing models, in future work, holographic radio will need a featured theory and modeling, converging the communication and electromagnetic theories. Moreover, performance estimation of communication requires dedicated electromagnetic numerical computation, such as the algorithms and tools related to computational electromagnetics and computer holography. The massive MIMO theory can be extended to make optimal use of these propagation models.

As mentioned above, a hierarchical and heterogeneous optoelectronic computing architecture is a key to holographic radio. Research challenges related to the hardware and physical layer design include the mapping from RF holography to optical holography, integration between photonics-based continuous-aperture active antennas and high-performance optical computing.\footnote{The authors would like to thank Danping~He\footref{bjtu} for the fruitful discussions on the topics of this section.}

%-------------------------------------------------------------------------
\subsection{Intelligent Reflecting Surfaces}

When the carrier frequency is increased, the wireless propagation conditions become more challenging due to the larger penetration losses and lower level of scattering, leading to fewer useful propagation paths between the transmitter and receiver. Moreover, designing coherently operating antenna arrays becomes more difficult since the size of each antenna element shrinks with the wavelength. In such situations, an intelligent reflecting surface (IRS) can be deployed and utilized to improve the propagation conditions by introducing additional scattering and, particularly, control the scattering characteristics to create passive beamforming towards the desired receivers for achieving high beamforming gain as well as suppressing co-channel interference \cite{Liaskos2018a,Wu2019a}. 
Ideally, an IRS can create a smart, programmable, and controllable wireless propagation environment, which brings new degrees of freedom to the optimization of wireless networks, in addition to the traditional transceiver design. 

\begin{figure}[t]
\centering
\includegraphics[scale=0.65]{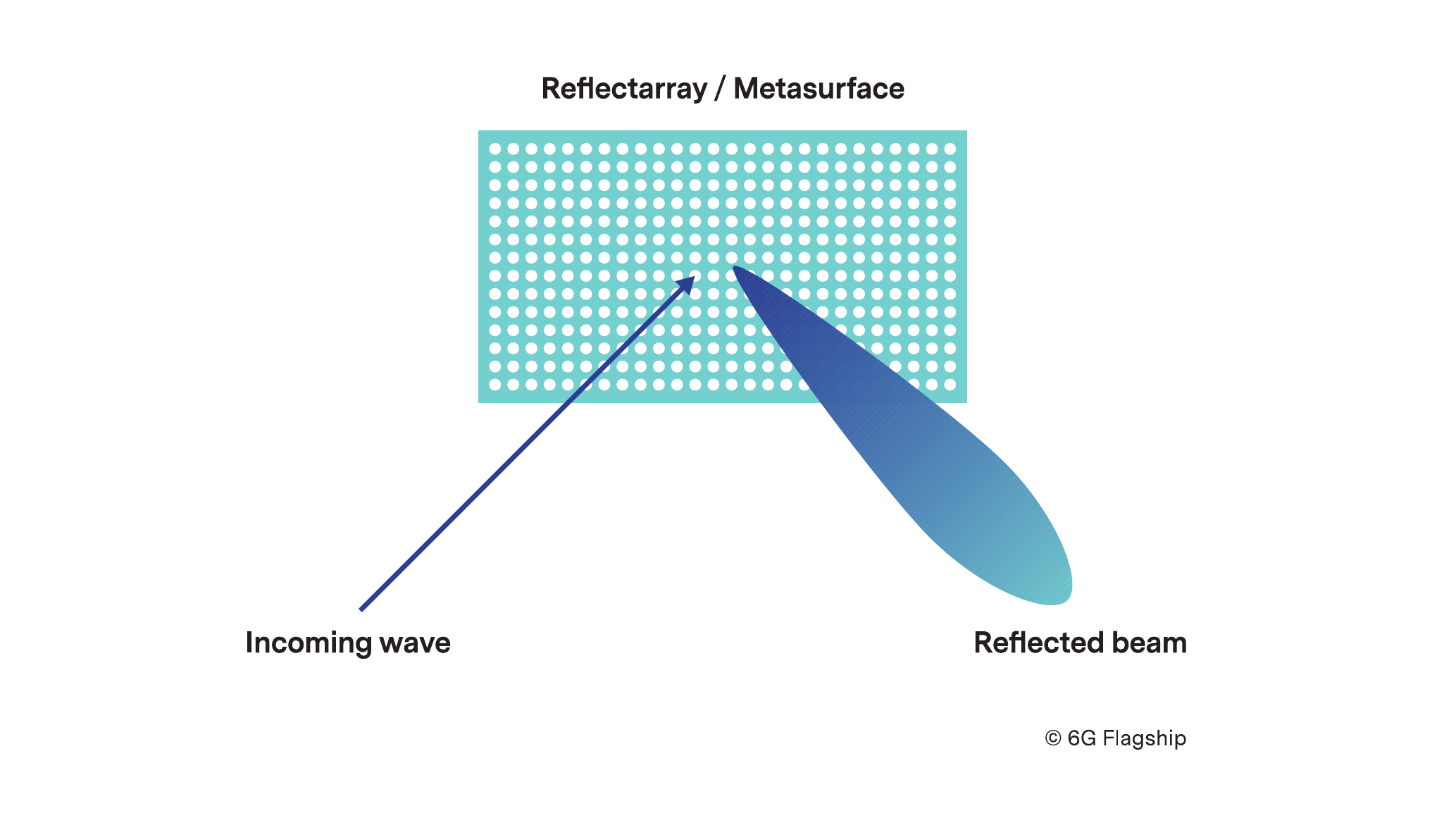}

\vspace{-6mm}

\caption{An IRS takes an incoming wave and reflects it as a beam in a particular direction or towards a spatial point.} \label{fig:metasurface}
\end{figure}

An IRS is a thin two-dimensional surface that can be implemented in different ways. The most capable option might be to make use of a metasurface consisting of metamaterials with unusual electromagnetic properties that can be controlled without the need for traditional RF chains. Large IRSs can be potentially produced with very low cost, complexity, and energy consumption since no RF components are required in general, compared to conventional active MIMO arrays. From an implementation standpoint, IRSs can be conveniently coated on facades of outdoor buildings or indoor walls/ceilings, making them deployable with low complexity and potentially invisible to the human eye. In addition, the easy practical fabrication of the IRS allows it to be mounted on arbitrarily shaped surfaces and can, therefore, be straightforwardly integrated into different application scenarios. Finally, the integration of an IRS into a wireless network can be made transparent to the users, thus providing high flexibility and superior compatibility.

The IRS concept has its origin in reflectarray antennas, which is a class of directive antennas that are flat but can be configured to act as parabolic reflectors or RF lenses \cite{Tretyakov2016,Headland2017}. A characteristic feature of an IRS is that it is real-time reconfigurable so it can be adapted to small-scale fading variations. The IRS contains a large number of sub-wavelength sized elements with highly controllable properties (e.g., impedance) that can be tuned to determine how an incoming signal is scattered; for example, the phase shift (i.e., time delay), amplitude and polarization. The consequence is that an incoming waveform can be reflected in the shape of a beam whose direction is determined by the phase-shift pattern over the elements and thus can be controlled \cite{Ozdogan2019a}. The reflection coefficients (e.g., phase-shifts), can be jointly optimized with the transmitters and receivers to maximize the performance such as the spectral efficiency or  energy efficiency of the end-to-end link \cite{Huang2018a}.
As illustrated in Figure~\ref{fig:metasurface}, it is not a matter of specular reflection but beamformed ``reflection'' in a wider meaning \cite{Bjornson2020IRS}. If the destination is in the vicinity of the IRS, the reflected signal can be focused on a particular spatial point. Since an IRS can change not only the direction of the reflected wave but also the shape of the waveform, it should be viewed as a reconfigurable lens rather than a mirror.

Another way to interpret an IRS is as a relay that does not amplify the incident signal but reflects it without causing any noticeable propagation delays (except that the delay spread of the channel might increase) \cite{JR:wu2018IRS}. Since the surface needs to be physically large to beat a classical half-duplex relay and is subject to beam-squinting \cite{Bjornson2020a,Bjornson2020IRS}, the most promising use case for IRS is to increase the propagation conditions in short-range communications, particularly in sub-THz and THz bands where relaying technology is not readily available. For example, the IRS can provide an additional strong propagation that remains even when the line-of-sight (LOS) path is blocked. An IRS should ideally be deployed to have a LOS path to either the transmitter and/or receiver. However, its use cases in non-LOS scenarios are also important, for example, to increase the rank of the channel to achieve the full multiplexing gain. 

In addition to increasing the signal strength of a single user, the surface can improve the channel rank (for both single-user and multi-user MIMO), suppress interference, and enhance multicasting performance. Essentially anything that can be done with traditional beamforming can also be implemented using an IRS, and they can be deployed as an add-on to many existing systems. Other prospective use cases are cognitive radio, wireless power transfer, physical layer security, and backscattering, where IoT devices near the surface can communicate with the AP at zero energy cost \cite{Liaskos2018a}. The most suitable use case is still an open question.

The main open research challenges are related to hardware implementation, channel modeling, experimental validation, and real-time control.
Although software-controllable metasurfaces exist \cite{Liaskos2018a}, the accuracy of the reconfigurability is limited, for example, by having a small number of possible phase values per element, fixed amplitudes for each phase-shift \cite{abeywickrama2019intelligent} and by only having control over groups of elements. Although theoretically, accurate channel models exist for LOS scenarios, these omit mutual coupling and other hardware effects that are inevitable in practice. Experimentally validated channel models are strongly needed. The performance loss caused by using practical low-resolution hardware such as 1-bit IRS also needs to be carefully studied, especially for the case when a large number of reflecting elements is required to achieve a considerable beamforming gain \cite{JR:wu2019discreteIRS}. Furthermore, wireless networks, in general, operate in broadband channels with frequency selectivity. As such, the phase-shift pattern of the IRS needs to strike a balance between the channels of different frequency sub-bands, which further complicates the joint active and passive beamforming optimization.
Finally, the control interface and channel estimation are difficult when using a passive surface that cannot send or receive pilot signals; thus, channel measurements can only be made from pilots sent by other devices and received at other locations. 
The radio resources required for channel estimation and control will grow linearly with the number of elements, and thus be huge in cases of interest, unless a clever design that utilizes experimentally validated channel models can be devised.
The power of the control interface will likely dominate the total energy consumption of an IRS since there are no power amplifiers or RF chains.

%-------------------------------------------------------------------------
\subsection{User-Centric and Scalable Cell-Free Networking}

\begin{figure}[t]
\centering
\includegraphics[scale=0.6]{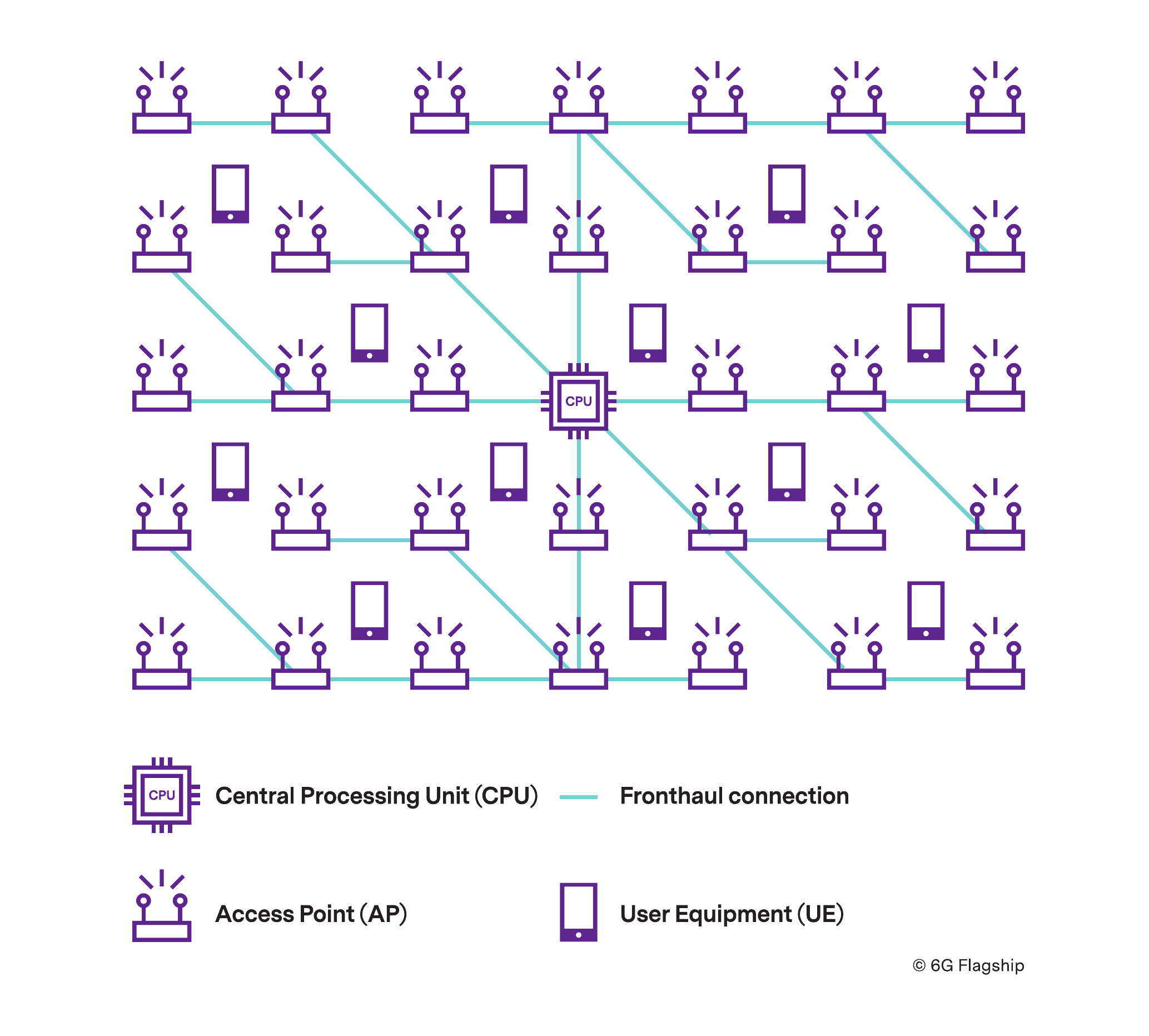}

\vspace{-6mm}

\caption{A Cell-free massive MIMO system consists of distributed APs that jointly serve the UEs. The cooperation is facilitated by a fronthaul network and a CPU.}
\label{fig:cell-free}
\end{figure}

The large variations in performance between cell center and cell edge is a main drawback of the conventional cellular networks. The concept of cell-free massive MIMO is a 6G technology designed for achieving nearly uniform performance and seamless handover across the user equipments (UEs) regardless of their position \cite{Ngo17,Nayebi2017a,BuzziUCletter}. By conveniently combining the best elements from massive MIMO, small cells, user-centric network MIMO, and coordinated multi-point (CoMP) with joint transmission/reception~\cite{Boldi2011a,Bjornson2013d,Int19,Liu18,Buz20}, it aims at creating a cell-free network where the achievable rates are nearly uniform across the coverage area. This is realized by replacing traditional APs, equipped with large co-located antenna arrays, with a large number of low-cost APs equipped with few antennas each. These APs are distributed over the coverage area and are cooperating via a fronthaul network and connected to one or multiple central processing units (CPUs). This architecture is illustrated in Figure~\ref{fig:cell-free}. Cell-free networks are ideally operated in time-division-duplex (TDD) mode so that uplink pilot signals can be utilized for both uplink and downlink channel estimation.

By distributing the antennas over a large number of APs, instead of using a smaller number of conventional APs, each UE is located, with large likelihood, at a short distance from at least one AP while also being able to satisfactorily communicate with a few other APs. As a consequence, even if each UE is no longer connected to an extremely large number of antennas (as in traditional massive MIMO), it is served by a reasonable number of ``good'' antennas with reduced pathloss. On top of this, since these antennas belong to spatially separated APs (which are usually seen by the UE with very different angles), large-scale fading diversity can be achieved, as opposed to the case of co-located antennas. Differently from the well-known concepts of CoMP and network MIMO, which are normally implemented in a network-centric fashion where edges between cooperation regions still exist,  cell-free massive MIMO can be realized by adopting a user-centric approach \cite{Bjornson2013d,Buz20,Bjornson2019}, whereby each UE is served only by the nearby APs.

The cell-free massive MIMO architecture is not meant for increasing the peak rates in broadband applications, since these can only be achieved in extreme cases, but has been shown to vastly outperform traditional small-cell and cellular massive MIMO for the majority of users \cite{Ngo17,Nayebi2017a,Bjornson2020a}. The cell-free massive MIMO deployment can also provide support for the implementation of low-latency, mission-critical applications. The availability of many APs, coupled with the rapidly decreasing cost of storage and computing capabilities, permits using cell-free massive MIMO deployments for the caching of content close to the UEs and for the realization of distributed computing architectures, which can be used to offload network-intensive computational tasks. Moreover, in low-demand situations, some APs can be partially switched off (control signals may still be transmitted) with a rather limited impact on the network performance, thus contributing to reducing the OPEX of mobile operators and their carbon footprint.

\subsubsection{Cell-Free Initial Access}

\begin{figure}[t!]
\centering
\includegraphics[scale=0.6]{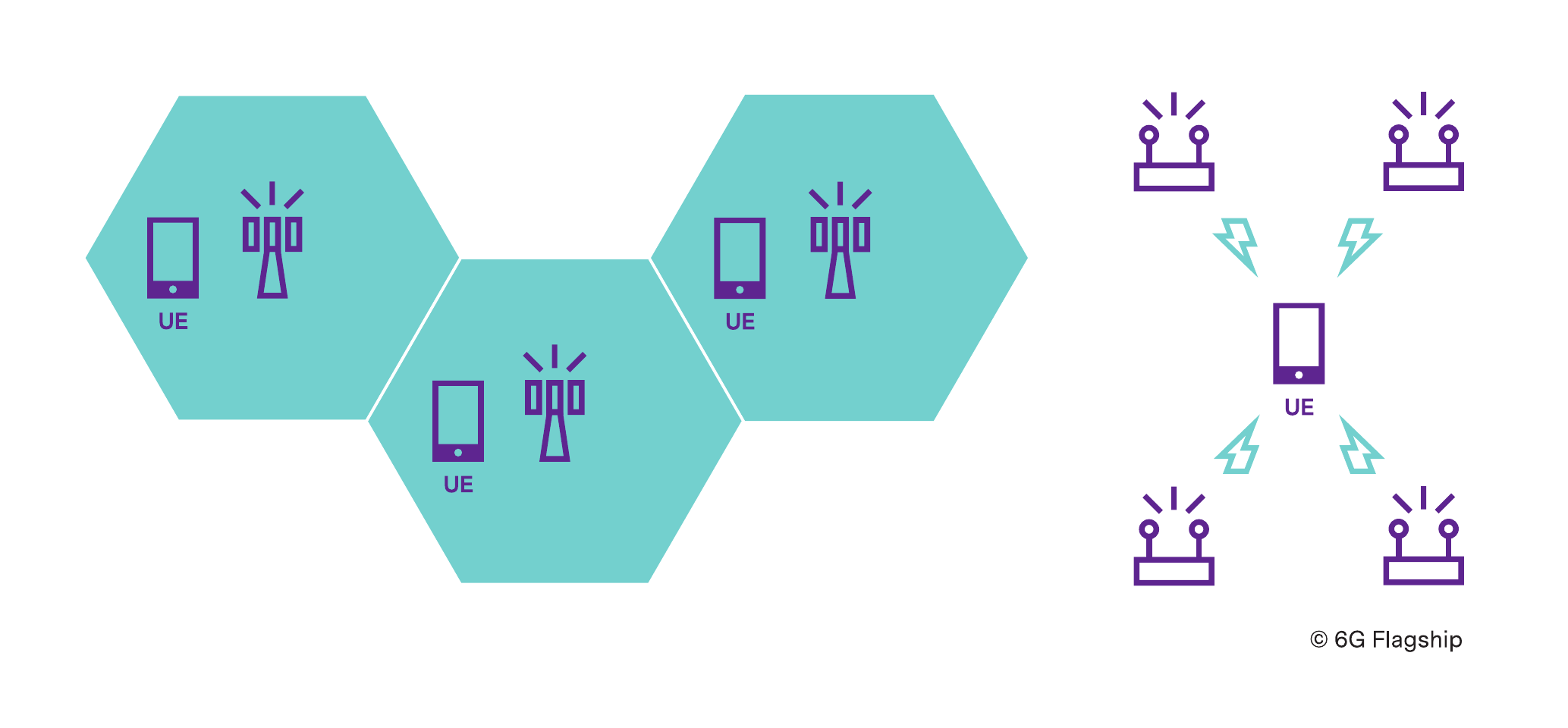}

\vspace{-6mm}

\caption{Initial access based on cellular system (left) and cell-free system (right).}
\label{fig:cfac}
\end{figure}

The basic connection procedures of cell search and random access determine the network performance in terms of latency, energy consumption, and the number of supported users \cite{3gpp300}. These basic functionalities are currently tailored to the cellular architecture. Figure~\ref{fig:cfac} illustrates initial access based on different mechanisms. As shown in the figure, in the cellular system, a UE is attempting to access a  cell in a cellular network. The UE usually chooses the strongest cell based on the measurement of received synchronization signals and will be subject to interference from neighboring cells. 
On the contrary, in the cell-free system, a UE is attempting to access a cell-free network where all the neighboring APs are supporting the UE's access to the network. To enable this, the traditional cell identification procedure must be re-defined, including the synchronization signals and how system information is broadcast.
Similarly, a new random access mechanism suitable for cell-free networks is needed such that some messages in the random access procedure can be transmitted and processed at multiple APs.

In summary, it is desirable that idle/inactive UEs can harness the benefits of the cell-free architecture as much as the active UEs can. To realize this, it is imperative to redesign the procedures of cell search and random access. To reduce the latency and improve the resource utilization efficiency, NOMA-enabled two-step RACH \cite{3gpp812} or autonomous grant-free data transmission \cite{yuan2018} should be investigated in cell-free networks.

\subsubsection{Implementation Challenges}

The huge amount of channel state information (CSI) that needs to be exchanged over the fronthaul to implement centralized joint precoding/combining is a long-standing scalability bottleneck for the practical implementation of CoMP and network MIMO~\cite{Bjornson2013d,Kal18}. In its original form \cite{Ngo17,Nayebi2017a}, cell-free massive MIMO avoids CSI exchange by only performing signal encoding/decoding at the CPU, while combining/precoding (such as matched filtering, local zero-forcing, and local minimum mean squared error (MMSE) processing) is implemented at each AP using locally acquired CSI. By synchronizing the APs, the signals can be coherently combined without the need for CSI sharing. In this case, each AP consists of antennas and UE-grade RF modules that perform digital operations such as channel estimation, combining/precoding, interpolation/decimation, digital pre-distortion, and DFT \cite{Int19}.

However, the performance of cell-free massive MIMO systems can be sensibly boosted by increasing the level of coordination among the APs, in particular, by sending the CSI to the CPU and perform joint combining/precoding \cite{Nayebi2016a,Bjornson2020a} as in conventional CoMP and network MIMO. In this case, the APs can be made smaller since most of the digital operations are carried out at the CPU. While the data rates can be increased, the drawback is higher computational complexity and latency, and the fronthaul signaling might also increase. If there are 64 antennas taking 20-bit samples, the total fronthaul capacity will far exceed 1 Gbps at the sampling rates of interest. This imposes significant challenges for interconnection technologies, such as Common Public Radio Interface (CPRI) and on the I/O interface of application-specific integrated circuits (ASIC) or field-programmable gate arrays (FPGA)~\cite{Li18}. A potential solution to this problem can be forming separate antenna clusters and having separate CPRI for each cluster. However, this increases the overall complexity of the system. Over-the-air bi-directional signaling between the APs and the UEs might be utilized as a flexible alternative to fronthaul signaling~\cite{Tol19,Atz20,Atz20a}. Furthermore, it is necessary to investigate scalable device coordination and synchronization methods to implement CSI and data exchange \cite{Bjornson2019}.

To provide the aforementioned gains over cellular technology,  cell-free massive MIMO requires the use of a large number of APs and the attendant deployment of suitable fronthaul links. Although there are concepts (e.g., radio stripes where the antennas are integrated into cables \cite{Int19}) to achieve practically convenient deployment, the technology is mainly of interest for crowded areas with a huge traffic demand or robustness requirements. The cell-free network will probably be underlaying a traditional cellular network, likely to use APs with large co-located arrays. The potentiality offered by integrated access and backhaul techniques can also be very helpful in alleviating the fronthaul problem and in reducing the cost of deployment.

Cell-free massive MIMO can be deployed in any frequency band, including below-6 GHz, mmWave, sub-THz, and THz bands. In the latter cases, the APs can serve each UE using a bandwidth of $100$~GHz or higher, which yields extremely high data rates over short distances and low mobility. The spatial diversity gains of the cell-free architecture become particularly evident in such scenarios because the signal from a single AP is easily blocked, but the risk that all neighboring APs are simultaneously blocked is vastly lower.\footnote{The authors would like to thank Tachporn~Sanguanpuak\footref{cwc}  and Shahriar~Shahabuddin\footref{nokia}  for the fruitful discussions on the topics of this section.}

%-------------------------------------------------------------------------
\subsection{Integrated Access and Backhaul}

At mmWave frequencies and above, there is a need for dense network deployments to mitigate the reduced coverage at these bands due to constraints in the propagation environment and hardware limitations in the transceivers. Such dense deployments make backhauling/fronthauling challenging at these bands, in particular, it is expensive and cumbersome to roll out fiber links to all APs. However, the presence of very wide bandwidths at these carrier frequencies makes it possible to include the wireless backhaul/fronthaul in the same spectrum as the wireless access. For this reason, integrated access and backhaul (IAB) network configurations seem promising, where a few (potentially, fiber-connected) APs provide other APs as well as the mobile devices inside their cell area with wireless backhaul and access connections, respectively \cite{massivemimobook,HCJ+2017}. IAB networks are different from conventional relay networks, because the information load changes in different hops. Particularly, as the number of mobile devices per hop increases, the APs need to transfer an aggregated data of multiple mobile devices accumulated from the previous hops. In addition, the interference on the links will depend on the information load. Some early work has been performed. It was demonstrated in \cite[Sec.~7.6]{massivemimobook} how massive MIMO can be used for backhaul to small cells. In this framework, the access and backhaul are separated in space. In \cite{SAD2018}, an alternative framework for an IAB-enabled cellular network was developed and used to characterize the downlink rate coverage probability. In this framework, the bandwidth was divided between access and backhaul, either statistically or based on the dynamic network load. Given the fact that 6G networks will see an even greater level of densification and spectrum heterogeneity than 5G networks, it is expected that IAB networks will play a major role, especially at upper mmWave and THz carrier frequencies.

%-------------------------------------------------------------------------
\subsection{Integrated Space and Terrestrial Networks}

\begin{figure}[t]
\centering
\includegraphics[scale=0.6]{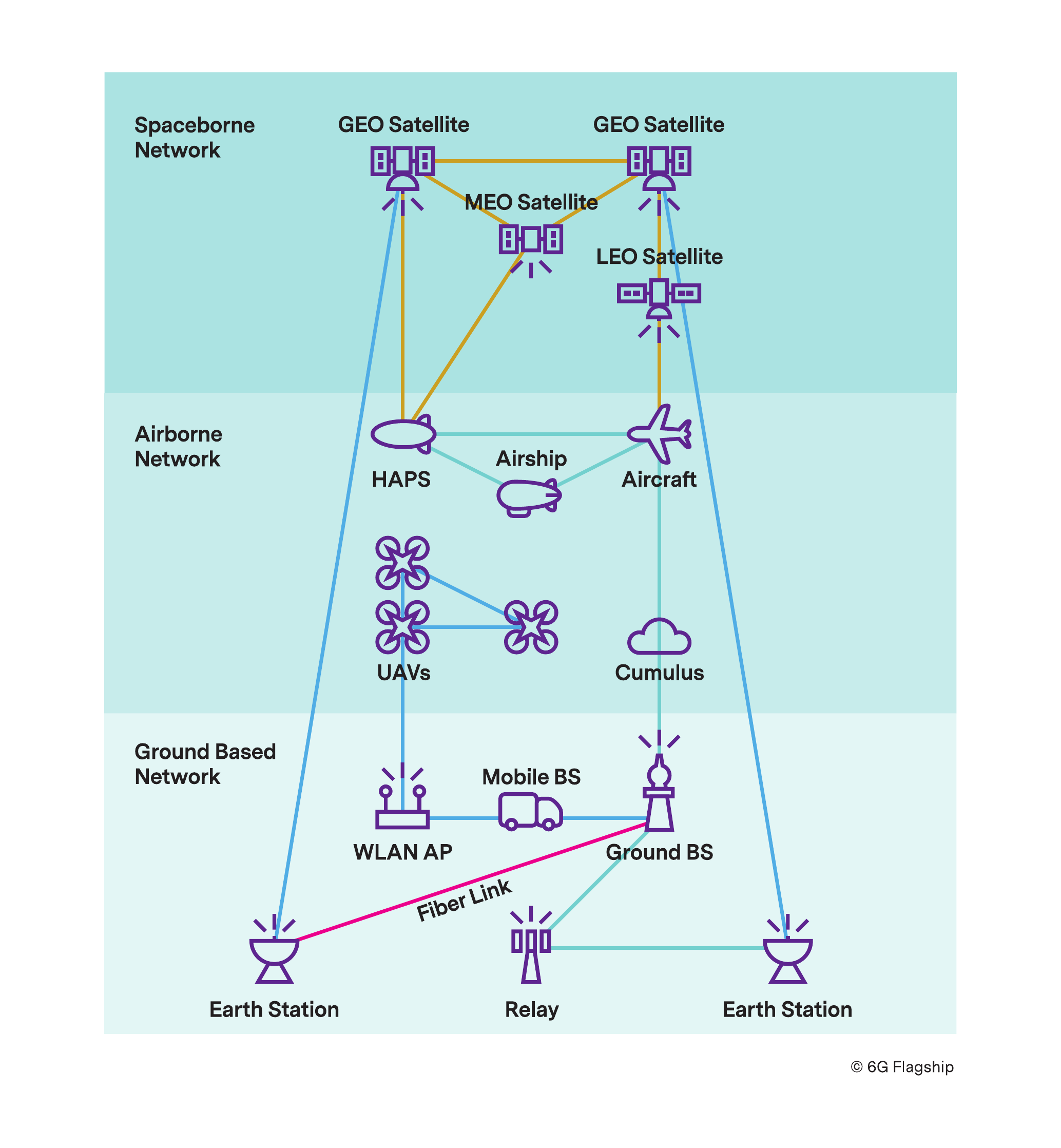}

\vspace{-8mm}

\caption{Illustration of the ISTN architecture and its three layers. Red line: free space optics link; green line: mmWave link. Blue line: other microwave link).}
\label{fig:ISTN}
\end{figure}

The 5G wireless systems are ground-based, having the same coverage problems as other terrestrial networks. Space communication networks are complementary to terrestrial networks as they provide outdoor communication coverage for people and vehicles at sea, as well as in remote rural areas and in the air. 6G networks should seamlessly integrate space networks with terrestrial networks to support truly global wireless communications anywhere and anytime \cite{HuangVTM2019}. An architectural illustration of a typical integrated space and terrestrial network (ISTN) is shown in Figure~\ref{fig:ISTN} and can be divided into three layers: the spaceborne network layer, the airborne network layer, and the conventional ground-based network layer. The spaceborne layer consists of LEO, medium Earth orbit (MEO), and geosynchronous orbit (GEO) satellites. The LEO category is particularly interesting for broadband applications since the shorter distance from the ground leads to better SNR and latency; in fact, the latency of long-distance transmission can be better than if cables are utilized on the ground. The drawback is that the many LEOs are needed to cover the Earth and synchronization is more complicated since they are constantly in motion.
Although the communication links are LOS-dominant, a main challenge for ISTN is that the large distances lead to relatively large propagation losses between the layers. Stationary high-gain antennas are traditionally used for communication between satellites and the ground, but lack the flexibility needed for mobile applications. Another grand challenge is co-channel interference, which is more severe in LOS applications. Hence, two predominant bottlenecks are the bandwidth available for aerial backbones and the limited area spectral efficiency of direct air-to-ground communications between the airborne and ground-based networks, both related to the airborne network.

As the airborne network is an intermediate layer between the spaceborne and the ground-based networks, high-speed, flexible, and all-weather aerial backbones constitute the most critical infrastructural elements in building the airborne network. A mmWave airborne link is an attractive option due to the wide bandwidths and relatively mature technology. This is indeed the spectrum currently used for communication between satellites and ground stations. Such high-speed links can be realized by using adaptive antenna arrays and high transmit power, possibly in conjunction with massive MIMO or in-band full-duplex operation. Since the topological structure of the airborne platform is in a constantly changing state, tracking the trajectory of the airborne platform in motion while supporting reliable continuous communications is also an open challenge. If the channel state information required for coherent spatial multiplexing is not available, then the combination of adaptive beamforming and space-time coding might be a suitable alternative.

While the ground-based and airborne networks can be physically managed by engineers, which can install software updates or replace faulty components when needed, the situation is totally different in the satellite layer. Due to the high cost of satellites and lack of physical access to them after being launched into space, only well-proven hardware and software can be used to guarantee the long-term operation. Hence, this is not a layer where the latest, innovative technology concepts should be used, but rather technologies that have a long track record. The processing and transmit power is also limited by the energy that the satellites can harvest from the sun. 

Maintaining reliable communications from/to the airborne platforms is of critical importance. With the adoption of mmWave spectrum for both the air-to-air and air-to-ground backbones as well as the direct air-to-ground access links, in-depth understanding of the aeronautical mmWave propagation characteristics is absolutely necessary. The aeronautical channel model for mmWave links is currently unavailable and significant efforts should be made to model it as a function of the carrier frequency, airborne platform velocity, Doppler frequency, delay spread, Rician K-factor, the number of multipath components of NLOS channels, and the array sizes, just to mention a few. There are many complex networking issues that must be studied, such as diverse types of communication links, various link delays, intermittent transmissions between satellites, dynamically fluctuating topology, high Doppler frequency, and multi-layer interference.

%-------------------------------------------------------------------------
\subsection{Integrated Broadcast and Multicast Networks} \label{section:broadcast_multicast}

The demands on network capabilities continue to increase in terms of capacity, availability and cost. These increasing demands can be challenging to networks that only support unicast, particularly, when there are more users that require simultaneous service than the APs is able to separate by beamforming. When some of those users are requesting the same data at the same time, broadcast and multicast are suitable transport mechanisms for large-scale delivery, since they permit to transmit the same content to a vast number of devices within the covered area at the same time and with a predefined quality-of-service.

Future 6G delivery networks need to be as flexible as possible to respond to the needs of service providers, network operators and users. Hence, broadcasting/multicasting could be integrated into 6G as a flexible and dynamic delivery option to enable a cost-efficient and scalable delivery of content and services in situations where the APs have limited beamforming capability. More precisely, if the AP can only send wide beams, then it should be able to broadcast information over the coverage area of that beam.

6G also represents a great opportunity for the convergence of mobile broadband and traditional broadcast networks, for example, used for TV broadcasting. Low-Power Low-Tower (LPLT) cellular networks with adaptive beamforming capabilities would benefit from the complementary coverage provided by High-Power High-Tower (HPHT) broadcast networks with fixed antenna patterns. A technology flexible enough to efficiently distribute content over any of these networks would allow the 6G infrastructure to better match the needs of future consumers and make more efficient use of existing tower infrastructure.

\begin{figure}[t]
\centering
\includegraphics[scale=0.6]{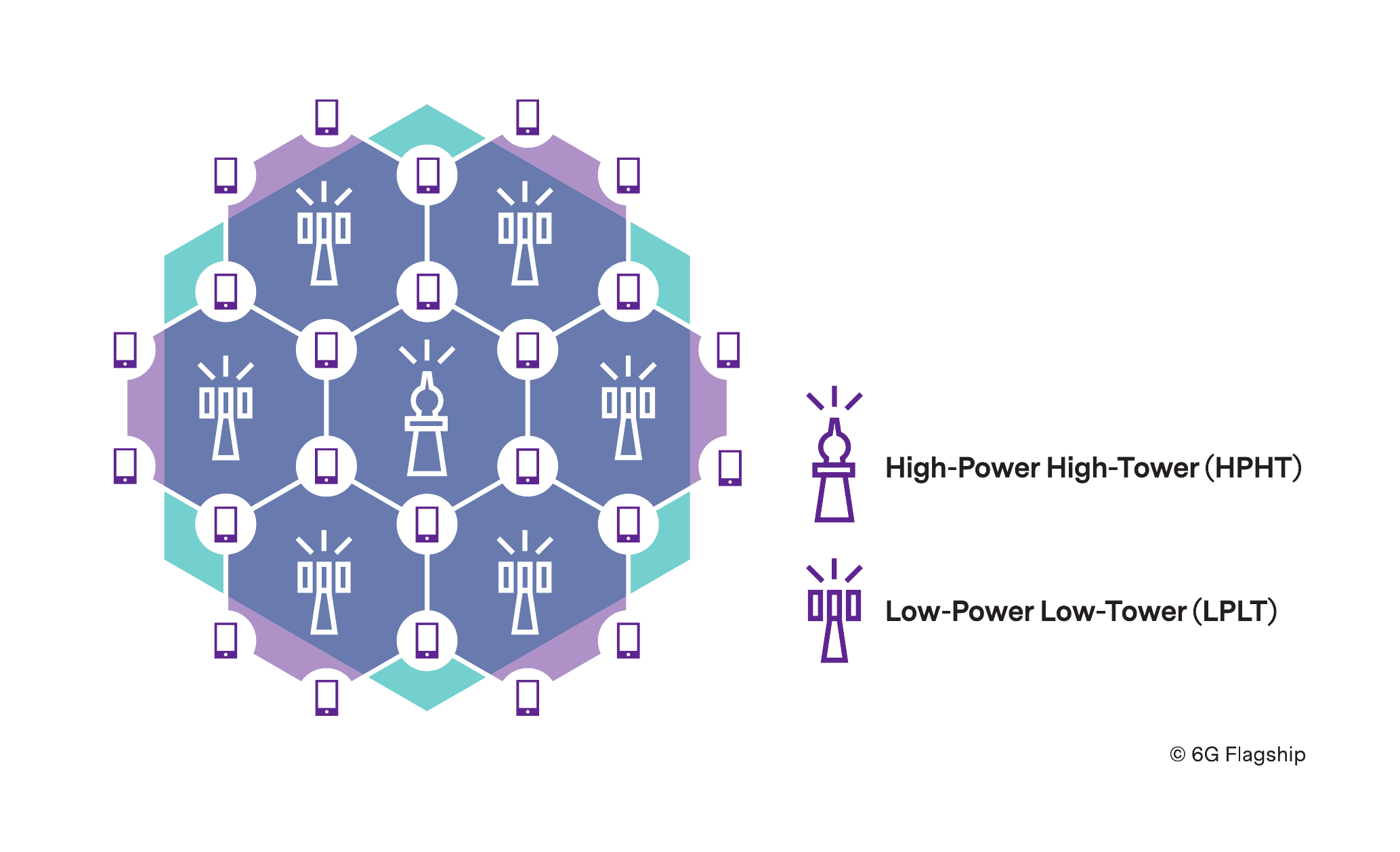}

\vspace{-8mm}

\caption{Network convergence and different topologies involved.}
\label{Figure_2}
\end{figure}

This convergence could be addressed from different perspectives; for example, the design of a single and highly efficient radio physical layer, the use of common transport protocols across fixed and mobile networks including broadcast, multicast, and unicast delivery, or a holistic approach that allows client applications running on handsets to better understand and, therefore, adapt to the capabilities of the underlying networks.

In Europe, it was decided to ``ensure availability at least until 2030 of the 470-694 MHz (‘sub-700 MHz’) frequency band for the terrestrial provision of broadcasting services'' \cite{EUJournal}. Some assignations are already happening in the USA, where the 600 MHz frequency band is assigned to mobile broadband services \cite{Gomez-Barquero}. The timing is very well aligned with the release of 6G, and it is at this moment when the convergence could take place. 

\begin{figure}[t]
\centering
\includegraphics[scale=0.6]{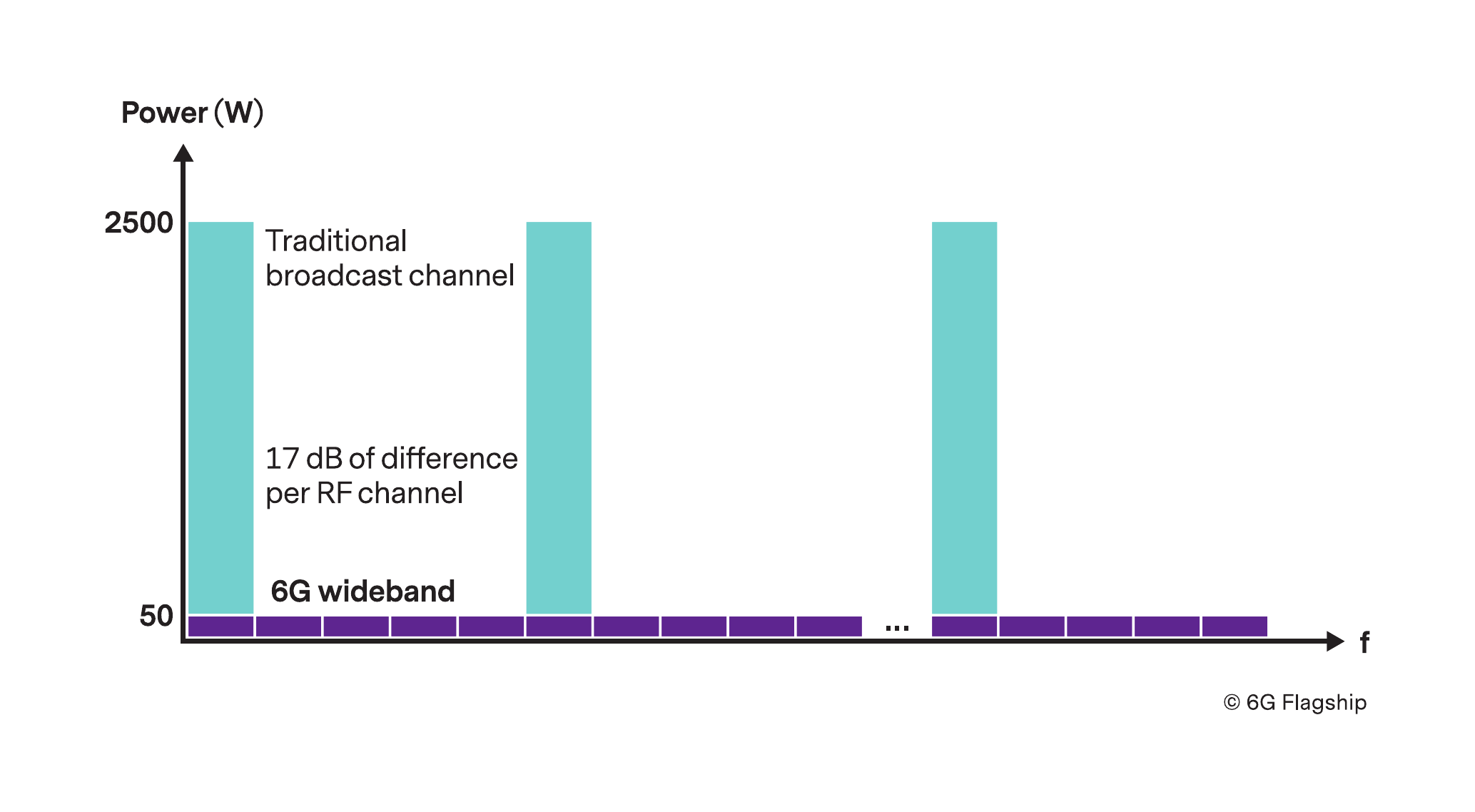}

\vspace{-8mm}
\caption{6G wideband concept and comparison with traditional digital broadcast. Source: \cite{Stare}}
\label{Figure_3}
\end{figure}

A potential solution would be the use of a 6G wideband broadcasting system  \cite{Stare}, where all RF channels within a particular frequency band are used on all HPHT transmitter sites (i.e., reuse-1). This is drastically different from current broadcasting networks, where usually a reuse-7 is used to avoid inter-cell interference. The entire wideband signal requires only half the transmission power of a single traditional digital RF channel, see Figure \ref{Figure_3}. By using 6G wideband, a similar capacity could be obtained with a much more robust modulation and coding rate, since the whole frequency band is employed. This represents the opportunity of transmitting about 17 dB less power (around 50 times) per RF channel (8 MHz), although using more RF channels per station, which leads to a total transmit power saving of around 90\%. Thanks to this higher spectrum use, the approach allows not only for a dramatic reduction in fundamental power/cost but also about 37-60\% capacity increase for the same coverage as with current services.

%=========================================================================
\section{Enablers at the Spectrum Level}
%=========================================================================

%-------------------------------------------------------------------------
\subsection{From mmWave to THz Communications}

\subsubsection{State of the Art of mmWave Systems (sub-100 GHz)}

The 3GPP has recently decided to start the study of Rel-17, including aspects for 5G NR operation beyond 52.6~GHz. The main interest will be to first extend the current NR frequency range 2 (FR2) support to the frequency range from 52.6~GHz to 71~GHz with minimal changes to the system. It is then expected that future studies will continue beyond 71~GHz towards the THz band, and the current physical layer optimized for below 52.6~GHz will then need to be significantly revised. One of the reasons for this is the fact that the transceiver impairments, such as phase noise (PN), will increase drastically. Another significant challenge is the reduction of the power amplifier efficiencies, which limits the coverage. To this end, the supported waveforms and subcarrier spacing (SCS) will be some of the first design aspects to be revised, because these will have a significant effect on the achievable bandwidth, impairment robustness, and coverage.

The currently supported waveforms in 5G NR are OFDM and DFT-spread-OFDM (or DFT-s-OFDM). The latter is a single-carrier (SC)-like transmission scheme mainly designed for coverage-limited cases due to its better power efficiency. It is currently only supported for uplink and rank-one transmissions. From the perspective of \emph{legacy systems}, it is always mandatory to first recognize the potential problems of the existing waveforms and then find solutions to improve the performance. It is known that DFT-s-OFDM can provide 3-5~dB better output power than OFDM, and the gain comes especially in the low-PAPR transmission schemes with low modulation orders \cite{6G2020Tervo}. This is one clear point that could make SC waveforms more efficient than OFDM, both in uplink and downlink, when going to very high frequencies. Moreover, SC waveforms are significantly more robust to PN because they can make use of the time-domain phase tracking reference signals (PTRS), which enables efficient PN estimation and compensation at the receiver.

In this direction, it has been shown that the current 5G NR DFT-s-OFDM can very well scale at least up to 90~GHz carrier frequencies for lower-order modulations (even up to 64-QAM) by optimizing the PTRS and PN estimation designs, which makes the DFT-s-OFDM or other SC waveforms promising candidates for even higher frequencies \cite{6G2020Tervo}. However, high-order modulations are more difficult to accommodate, and the use of those may require to increase SCS to reduce the impact of PN. Thus, one good way to achieve good spectral efficiency and coverage could be to use low-order modulation (which has low PAPR and can work with quite severe PN) with low-order spatial multiplexing. Another reason for higher required SCS is to enable higher bandwidths with reasonable FFT size. 

While the current 3GPP studies are only considering spectrum below 100 GHz, 6G is also expected to consider much higher bands and, thus, potentially much wider bandwidths.

\subsubsection{Ultra-Broadband Systems in the (Sub-)Terahertz Band (Above 100 GHz)}

Terahertz-band communication~\cite{akyildiz2014teranets,kurner2014towards,rappaport2019wireless} is envisioned as a 6G technology able to simultaneously support higher data rates (up to around 1 Tbps) and denser networks (towards of billions of interconnected devices). For many years, the lack of compact, energy-efficient device technologies which are able to generate, modulate, detect, and demodulate THz signals has limited the feasibility of utilizing this frequency range for communications. However, major progress in electronic, photonic, and innovative plasmonic device technologies is finally closing the THz gap.

In parallel to device technology developments, many of the works have been focused on i) understanding the THz channel in LOS~\cite{jornet2011channel} and non-LOS (NLOS)~\cite{han2015multi} conditions, both in static~\cite{Guan2019Measurement} and mobile~\cite{guan2016millimeter} scenarios; ii) developing new modulation and coding schemes, including sub-picosecond-long pulse-based modulations able to maximize the utilization of the THz band over short distances~\cite{jornet2014femtosecond} as well as multi-carrier communication schemes~\cite{han2015multiw} for longer links; and, iii) devising strategies to increase the communication distance, including ultra-massive MIMO communication techniques able to support spatial multiplexing of 1024 signals~\cite{akyildiz2016realizing, sarieddeen2019terahertz}.

The main properties of the THz band channel that define the challenges and opportunities for 6G are as follows. At THz frequencies, the absorption by water vapor molecules defines multiple transmission windows, tens to hundreds of GHz wide each, within which the atmospheric losses are negligible (i.e., less than 2-3 dB). The position of such windows is fixed, but the actual usable bandwidth changes with the humidity content and the transmission distance. The second aspect to take into account is that the very small wavelength (sub-mm) of THz signals leads to antennas with a very small effective area and, thus, high-gain directional antenna designs are needed to compensate. Such antennas can be fixed antenna designs or antenna arrays with thousands of elements thanks to, again, the very small size of low-gain THz antennas.

\paragraph{First Stop: The Sub-THz Band, from 100~GHz to 300~GHz}~\\
In the quest for spectrum, it is expected that the \emph{first stop} will be in the sub-THz band, between 100 and 300~GHz, where technology is more mature. Recent measurement campaigns have shown that sub-THz propagation channels are largely dominated by the LOS path, which provides most of the energy contribution. It follows that the use of SC communication systems is envisaged to provide low-complexity RF and power amplifier (PA) architectures. In particular, sub-THz systems suffer from the strong phase impairments, PN and carrier frequency offset (CFO), resulting from the poor performance of high-frequency oscillators. Phase impairments actually deteriorate the communication performance in sub-THz bands severely. New device implementations, signal processing and communication solutions need to be developed to overcome these limitations.
For example, optimized modulation and demodulation schemes can be used to achieve some PN robustness~\cite{dore,majed}. Similarly, the design of interference cancellation algorithms is an ongoing research topic to alleviate the impact of CFO. Last, delivering high power (and relatively high efficiency) with CMOS compatible technologies is still an open research topic. Alternatively, the use of receivers based on envelope or energy detection is also considered, which enables a frequency down-conversion from passband to baseband without the impact of phase impairments. For instance, MIMO systems with energy detection receivers could offer a valuable solution to achieve high spectral efficiency sub-THz communications with low power and low complexity RF architectures. 

\begin{figure}[t]
\centering
\includegraphics[scale=0.75]{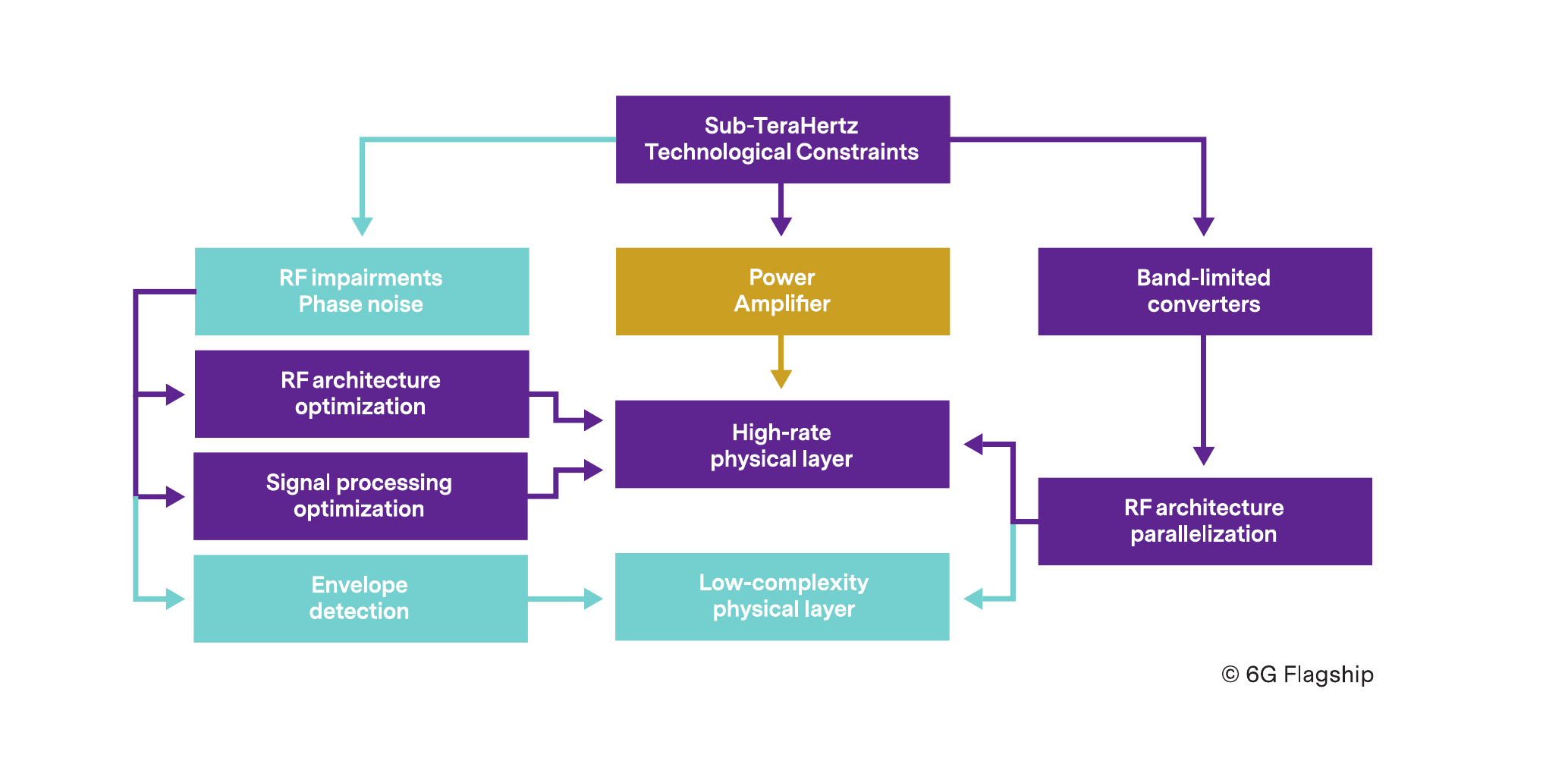}

\vspace{-8mm}

\caption{Illustration of the physical layer and RF paradigms for sub-THz communication systems.}
\label{fig:subthz}
\end{figure}

In light of this discussion, two paradigms arise for the physical layer (see Figure~\ref{fig:subthz}):

\begin{itemize}
\item \textbf{High-rate physical layer:} The high-rate physical layer corresponds to a communication system whose objective is to maximize the data rate for a given band, i.e., the spectral efficiency. The corresponding application is a high-capacity backhaul. This physical layer implies the use of in-phase/quadrature (IQ) transceivers, high-quality RF components, and high-order modulation schemes. Concerning the research on the physical layer for such communication systems, most of the current works investigate the optimization of channel bonding systems and the related signal processing.

\item \textbf{Low-complexity physical layer:} Conversely, in this paradigm, communication systems aim to minimize the complexity of the architecture or the energy consumption to achieve a given rate. Contemplated applications, in this case, are either the enhanced hotspot or short-range communications. This paradigm entails a complexity/power-limited regime and hence the use of simple RF architecture, analog or basic (e.g., on-off keying) modulation schemes. Numerous research approaches are under investigation for the development of the physical layer for complexity/energy-constrained systems. Research works include the use of index modulation, the design of high-rate impulse radio, the joint optimization of analog and digital signal processing, etc.
\end{itemize}

It is relevant to note that there is already an active THz communication standardization group, IEEE 802.15 IGTHz, which lead to the first standard IEEE 802.15.3d-2017.

\paragraph{Destination: Ultra-broadband Ultra-massive Systems at \emph{True} THz Frequencies}~\\
Enabling cellular systems at \emph{true} THz frequencies, in the infrared spectrum from 300 GHz and upwards, introduces further challenges. On the one hand, the much higher spreading losses (resulting from the much smaller wavelength of THz signals) combined with the low power of THz transmitters makes increasing the communication distance \emph{the grand challenge}~\cite{akyildiz2018combating}. Despite their quasi-optical propagation traits, THz communications possess several microwave characteristics. For instance, they can still make use of IRSs to support NLOS propagation~\cite{nie2019intelligent, ma2020intelligent, singh2020operation} and ultra-massive MIMO antenna array processing techniques to enable efficient beamforming~\cite{lin2016terahertz,yan2019dynamic,sarieddeen2019terahertz}. Although the spectrum ranges supported by digital beamforming will increase over time, less efficient hybrid or analog methods are likely to dominate. 
Adaptive arrays-of-subarrays antenna architectures in which each subarray undergoes independent analog beamforming is the first path to explore. New opportunities, including fully digital beamforming architectures with thousands of parallel channels~\cite{singh2020operation_tx}, might become a reality with new plasmonic device technologies. 
On the other hand, at higher frequencies, molecular absorption has a higher impact. As already explained, the absorption defines multiple transmission windows, tens to hundreds of GHz wide each. As a result, simple SC modulations can already enable very high-speed transmissions, exceeding tens of Gbps. Beyond the traditional schemes, dynamic-bandwidth algorithms that can cope with the distance-dependent absorption-defined transmission channel bandwidth have been proposed for short~\cite{han2016distance} and long~\cite{hossain2019hierarchical} communication distances. Ultimately, resource allocation strategies to jointly orchestrate frequency, bandwidth and antenna resources need to be developed.

An additional challenge in making the most of the THz band is related to the digitalization of large-bandwidth signals. While the THz-band channel supports bandwidth in excess of 100~GHz, the sampling frequency of state-of-the-art digital-to-analog and analog-to-digital converters is in the order of 100 Gigasamples-per-second. Therefore, high-parallelized systems and efficient signal processing are needed to make the most out of the THz band. Since channel coding is the most computationally demanding component of the baseband chain, efficient coding schemes need to be developed for Tbps operations. Nevertheless, the complete chain should be efficient and parallelizable. Therefore, algorithm and architecture co-optimization of channel estimation, channel coding, and data detection is required. The baseband complexity can further be reduced by using low-resolution digital-to-analog conversion systems, where all-analog solutions are also being considered. 

Beyond the physical layer, new link and network layer strategies for ultra-directional THz links are needed. Indeed, the necessity for very highly directional antennas (or antenna arrays) simultaneously at the transmitter and at the receiver to close a link introduces many challenges and requires a revision of common channel access strategies, cell and user discovery, and even relaying and collaborative networks. For example, receiver-initiated channel access policies based on polling from the receiver, as opposed to transmitter-led channel contention, have been recently proposed~\cite{xia2019link}. Similarly, innovative strategies that leverage the full antenna radiation pattern to expedite the neighbor discovery process have been experimentally demonstrated~\cite{xia2019expedited}.
All these aspects become more challenging for some of the specific use cases defined in Section~\ref{sec:use_cases}, for example, in the case of wireless backhaul, for which very long distances lead to very high gain directional antennas and, thus, ultra-narrow beamwidths, or smart rail mobility, where ultra-fast data-transfers can aid the intermittent connectivity in train-to-infrastructure scenarios.\footnote{The authors would like to thank Simon~Bicais\footref{cea}  for the fruitful discussions on the topics of this section.}

%-------------------------------------------------------------------------
\subsection{Optical Wireless Communications}

Beyond the THz spectrum, optical wireless communication (OWC) systems, which operate in the infrared (187-400~THz or, equivalently, from 1600~nm down to 750~nm wavelength), visible (400-770~THz / 390-750~nm) and even ultraviolet (1000-1500~THz / 200-280~nm) electromagnetic spectrum bands, are similarly being explored as a way to provide broadband connectivity. Optical wireless communications offer attractive features such as extremely high bandwidth, robustness to electromagnetic interference, a high degree of spatial confinement, inherent security, and unlicensed spectrum.

The term free space optical (FSO) communication is utilized to refer to OWC systems in the infrared frequency range. Such systems are commonly found at the basis of long-range high-speed point-to-point links utilized, for example, in ultra-broadband wireless backhaul applications~\cite{ciaramella20091} and, to a lesser extent, for indoor communications~\cite{gomez2014beyond}.
Here, we focus on visible light communication (VLC), or OWC in the visible light spectrum, which is a promising technology to provide broadband local connectivity~\cite{pathak2015visible}.
In VLC, all the baseband processing at the transmitter and the receiver is performed in the electrical domain and light-emitting diodes (LEDs) with a large field of view (FoV) or laser diodes (LDs) with small FoV are used to encode and transmit data over the LOS/NLOS optical channel. Photodetectors at the receiver convert data-carrying light intensity back to electrical signals for baseband processing.  

Performance-wise, data throughput below 100 Mbps can be achieved with relatively simple optical transceivers and off-the-shelf components. Data rates of up to hundreds of Gbps have been demonstrated in laboratory conditions. Key optical components for VLC, such as LEDs and photodetectors, have been developed for decades, and they are typically low-cost standard components. 
VLC is not intended to replace but complement existing technologies. When it comes to use cases, VLC can be used for both conventional data services for consumers and support emerging applications and services such as smart city, smart buildings, factories of future, intelligent transportation systems (ITS), smart grid and the internet of things (IoT). The concept of light-based IoT (LIoT) exploits light not only to create optical links but also to harvest its energy  \cite{Katz2019,Katz2020}. Thus, a LIoT node can be energy autonomous. VLC will also be useful in scenarios in which traditional RF communication is less effective such as in-cabin internet service in airplanes, underwater communication, healthcare zones, etc.

There are many practical constraints. For VLC systems, intensity modulation/direct detection (IM/DD) is the most practical scheme. The intensity-modulating data signal must satisfy a positive-valued amplitude constraint. Hence, it is not possible to straightforwardly apply techniques used in RF communications. A VLC-enabled device inside a pocket or briefcase cannot be connected optically. A solution to this would be a hybrid optical-radio wireless network. A reconfigurable optical-radio network is a high-performance and highly flexible communications system that can be adapted for changing situations and different scenarios \cite{Katz2020,Saud2019}.

%\begin{figure}[t!]
%\centering
%\includegraphics[width=14cm, keepaspectratio=true]{./Images/optical.png}
%\caption{6G vision: connectivity to be provided by radio and optical wireless communications.}
%\label{fig:optical}
%\end{figure}

%Figure~\ref{fig:optical} depicts a high-level vision of 6G, where wireless connectivity is provided by radio communications as well as optical communications to cover both local and wide area access. This is a clear departure from previous generations where radio was exclusively exploited. It is expected that the highly complementary nature of radio and optical wireless communications will be greatly exploited in 6G.

Open research directions for VLC systems include:
\begin{itemize}
\item Accurate VLC channel modeling and characterization for various deployment scenarios with a particular emphasis on user-dense environments. Incorporating user mobility and device orientation into the VLC channel models and combining VLC and RF systems, \cite{Miramirkhani2017,Uysal2017}
\item New non-coherent physical layer transmission schemes such as spatial modulation, and its variations, can be used, as well as other optical communications such as MIMO with non-orthogonal multiple access (NOMA) \cite{Basar2016,Haas2015,Yesilkaya2017}.
\item Exploiting RGB LEDs, development of new materials and optoelectronic devices (e.g., fast non-phosphorous LEDs, micro-LEDs), very fast switching mechanisms between optical and radio systems, etc.\footnote{The authors would like to thank Harald~Haas (Institute for Digital Communications, University of Edinburgh, UK) for the fruitful discussions on the topics of this section.}
\end{itemize}

%=========================================================================
\section{Enablers at the Protocol/Algorithmic Level}
%=========================================================================

%-------------------------------------------------------------------------
\subsection{Coding, Modulation, Waveform, and Duplex}

\subsubsection{Channel Coding}
From 4G to 5G, the peak data rate has increased by 10-100 times, and this trend is likely to continue with 6G. The throughput of a single decoder in a 6G device will reach hundreds of Gbps. Infrastructure links are even more demanding since they aggregate user throughput in a given cell or virtual cell, which is expected to increase due to spatial multiplexing. It is difficult to achieve such a high throughput, only relying on the progress of integrated circuit manufacturing technology within ten years. Solutions must be found from the algorithm side as well. Both code design and corresponding encoding/decoding algorithms need to be taken into account to reduce the decoding iterations and improve the decoder’s parallelism level. Moreover, it is vital for the decoder to achieve reasonably high energy efficiency. To maintain the same energy consumption as in current devices, the energy consumption per bit needs to be reduced by 1-2 orders of magnitude. Implementation considerations such as area efficiency (in Gbps/mm$^2$), energy efficiency (in Tb/J), and absolute power consumption (W) put huge challenges on code design, decoder architecture, and implementation~\cite{EPIC_project}.

6G communication systems require flexibility in codeword length and coding rate. The most commonly used coding schemes today are Turbo codes, Polar codes, and low-density parity check codes (LDPC). Their performance has already been pushed towards the limit using 16 nm and 7 nm technology \cite{EPIC_project}. Trade-offs must be made between parallelization, pipelining, iterations, and unrolling, linking with the code design and decoder architecture.
The future performance and efficiency improvements could come from CMOS scaling, but the picture is quite complex \cite{Kestel_2018}. Indeed, trade-offs must be made to cope with issues such as power density/dark silicon, interconnect delays, variability, and reliability. Cost considerations render the picture even more complex: cost due to silicon area and due to manufacturing masks, which explode at 7 nm and below.

The channel coding scheme used in 6G high-reliability scenarios must provide a lower error floor and better ``waterfall'' performance than that in 5G. Short and moderate length codes with excellent performance need to be considered.
Polar codes, due to their error correction capabilities and lack of error floor, might be the preferred choice in 6G. However, state-of-the-art CRC-aided successive cancellation list decoding doesn't scale up well with throughput due to the serial nature of the algorithm. Hence, iterative algorithms that can be parallelized need to be developed \cite{Elkelesh2018}. In \cite{Vism2020}, a new variant of the multi-trellis BP decoder is proposed, which permutes only a subgraph of the original factor graph. This enables the decoder to retain information of variable nodes in the subgraphs, which are not permuted, reducing the required number of iterations needed in-between the permutations. As a result, the proposed decoder can perform permutations more frequently, hence being more effective in mitigating the effect of cycles, which cause oscillation errors. Work in \cite{Heshani2019} proposes some new algorithms to process new types of node patterns that appear within multiple levels of pruned sub-trees, and it enables to process certain nodes in parallel. Furthermore, modified polar code constructions can be adopted to improve the performance of iterative algorithms, potentially with the aid of machine learning such as deep unfolding \cite{Gruber_2017,Nachmani2018}.
These approaches could lead to better communication performance but will require significant advances in understanding the behavior, robustness, and generalization of neural networks.

\subsubsection{Modulation and Waveform}

Modulation is another aspect that can be revised in 6G. High-order QAM constellations have been used to improve spectral efficiency in high SNR situations. However, because of the non-linearity of hardware, the benefits obtained in higher-order QAM constellations are gradually disappearing. 
New modulation methods, for example, the schemes based on signal shaping, have been adopted in the ATSC 3.0 standard and proved to be effective in optical fiber communication. Their application to wireless communication is worth careful study \cite{bocherer2015bandwidth,icscan2019probabilistic}. 

Reducing the peak-to-average-power-ratio (PAPR) is another important technology direction in order to enable IoT with low-cost devices, edge coverage in THz communications, industrial-IoT applications with high reliability, etc.
There will be many different types of demanding use cases in 6G, each having its own requirements. No single waveform solution will address the requirements of all scenarios. For example, as discussed also in Section 4.1, the high-frequency scenario is faced with challenges such as higher phase noise, larger propagation losses, and lower power amplifier efficiency. Single-carrier waveforms might be preferable over conventional multi-carrier waveforms to overcome these challenges \cite{OFDMvsSC}. 
For indoor hotspots, the requirements are instead the higher data rates and the need for flexible user scheduling. Waveforms based on OFDM or on its variants exhibiting lower out-of-band emissions \cite{UFMC1,UFMC2,FBMC} will remain a good option for this scenario. 
6G needs a high level of reconfigurability to become optimized towards different use cases at different time or frequency.

\subsubsection{Full-Duplex}

The current wireless systems (e.g., 4G, 5G, WiFi) are using time division duplex (TDD) or frequency division duplex (FDD), the so-called half-duplex, where transmission and reception are not performed at the same time or frequency. On the contrary, the full-duplex or the in-band full-duplex (IBFD) technology allows a device to transmit and receive simultaneously in the same frequency band. Full-duplex technology has the potential to double the spectral efficiency and significantly increase the throughput of wireless communication systems and networks. The biggest hurdle in the implementation of IBFD technology is self-interference, i.e., the interference generated by the transmit signal to the received signal, which can typically be over 100 dB higher than the receiver noise floor. Three classes of cancellation techniques are usually used to cope with the self-interference: Passive suppression, analog cancellation, and digital cancellation (see e.g. \cite{sabharwal2014in} and the references therein). Passive suppression involves achieving high isolation between transmit and receive antennas before analog or digital cancellation. Analog cancellation in the RF domain is compulsory to avoid saturating the receiving chain, and the nonlinear digital cancellation in baseband is needed to further suppress the self-interference, for example, down to the level of the noise floor. 

The full-duplex technique has a wide range of benefits, e.g., for relaying, bidirectional communication, cooperative transmission in heterogeneous networks, and cognitive radio applications. Its feasibility has been experimentally demonstrated in small-scale wireless communications environments, and it was also considered as an enabling technique for 5G but not yet adopted by 3GPP \cite{kolodziej2019in}. However, for the full-duplex technique to be successfully employed in 6G wireless systems, there exist challenges at all layers, ranging from the antenna and circuit design (e.g., due to hardware imperfection and nonlinearity, the non-ideal frequency response of the circuits, phase noise, etc), to the development of theoretical foundations for wireless networks with IBFD terminals. Note that IBFD becomes very challenging when MIMO is taken into account and is even more so with massive MIMO. The suitability of the IBFD technology for 6G is an open research area, where an inter-disciplinary approach will be essential to meet the numerous challenges ahead \cite{sabharwal2014in}.

%\begin{figure}[h]
%\includegraphics[width=12cm]{./Images/waveform}
%\centering \label{waveforms}
%\caption{Illustrations of 6G scenarios and suitable waveforms}
%\end{figure}

%-------------------------------------------------------------------------
\subsection{Interference Management Using NOMA and Rate-Splitting}

The purpose of cellular networks is to provide wireless access to multiple users, which share the radio resources in a controllable manner to achieve satisfactory quality-of-service. The traditional solution was to provide orthogonal multiple access within each cell, while the inter-cell interference was non-orthogonal but controlled using frequency reuse patterns. This situation changed with 4G, where co-user interference was instead managed in the spatial domain using sectorization and beamforming. In 5G,
massive MIMO is used to control co-user interference at the physical layer, by adaptive beamforming, and many spatially separated users can thereby be served non-orthogonally by spatial multiplexing.

However, there are situations when the interference management provided by massive MIMO is insufficient, for example, if the spatial resolution of the antenna panels is too limited \cite{Senel2019}, which can happen when the array is far from the users.
In these cases, interference can also be managed at the protocol level, using so-called non-orthogonal multiple access (NOMA) schemes \cite{Liu2017} or rate-splitting (RS) \cite{Clerckx2016a}. The NOMA techniques are mainly based on two domains: power and code \cite{Dai2018}. Power-domain NOMA refers to the superposition of multiple messages using different transmit power, such that users with higher SNRs can decode interfering signals before decoding their own signal, while users with lower SNRs can treat interference as noise. 
Code-domain NOMA refers to the use of non-orthogonal spreading codes, which provide the users with higher SNRs after despreading, at the expense of additional interference.
RS is based on dividing the users' messages into private and common parts. Each user decodes its private part and the common parts to extract its data. RS can be viewed as a generalization of power/code-domain NOMA.

There are several practical challenges with implementing robust interference management at the protocol level. One is the error propagation effects that appear when applying superposition coding and successive interference cancellation to finite-sized data blocks. Another issue is the high complexity in jointly selecting the coding rates of all messages, to enable successful decoding wherever needed, and conveying this information to the users. Implementing adaptive modulation and coding is nontrivial in fading environments with time-varying interference and beamforming, even if it is carried out on a per-user basis. With NOMA and RS, the selection of modulation/coding becomes coupled between the users, making it an extremely challenging resource allocation problem. Scheduling and ARQ retransmissions are other tasks that become more complex when the user data transmission is coupled. Hence, there is a large need to study protocol-level interference management schemes under practical conditions.

The non-orthogonal access and interference management provided by massive MIMO were considered sufficient in 5G; thus, neither NOMA nor RS was included in the first releases. However, there are prospective 6G use cases where interference management at the protocol level can potentially provide sufficiently large gains to outweigh the increased implementation complexity.
In massive connectivity scenarios, where many devices transmit small packages intermittently, grant-free access using code-domain NOMA or similar protocols can be very competitive.
In mmWave or THz communications where the controllability of the beamforming is limited by hardware constraints (e.g., phased arrays), NOMA and RS can enable multiple users to share the same beam \cite{Zhu2019}. There is also a need for multiple access schemes in VLC, where coherent adaptive beamforming might be practically impossible. NOMA and RS might be the required solution \cite{Marshoud2016a}.

\subsection{Machine Learning-Aided Algorithms}

The wireless technology becomes more complicated for every generation with many sophisticated interconnected components to design and optimize. In recent research, there have been an increasing interest in utilizing machine learning (ML) techniques to complement the traditional model-driven algorithmic design with data-driven approaches. There are two main motivations for this \cite{Simeone2018a}: modeling or algorithmic deficiencies.

The traditional model-driven approach is optimal when the underlying models are accurate. This is often the case in the physical layer, but not always. When the system operates in complex propagation environments that have distinct but unknown channel properties, there is a modeling deficiency and, therefore, a potential benefit of tweaking the physical layer using ML.

There are also situations where the data-driven approach leads to highly complex algorithms, both when it comes to computations, run-time, and the acquisition of the necessary side information. This is called an algorithmic deficiency and it can potentially be overcome by ML techniques, which can learn how to take shortcuts in the algorithmic design. This is particularly important to obtain effective signal processing for latency-critical applications and when jointly optimizing many blocks in a communication system.

This section provides some concrete examples and potential future research directions.

\subsubsection{End-to-End Learning}

End-to-end optimization of a communication system using ML is introduced in \cite{8054694_autoencoder}, where the system is interpreted as an autoencoder that jointly optimizes the transmitter and receiver components in a single process to minimize the end-to-end reconstruction error. This concept is mainly of interest when the channel model is unknown or difficult to model analytically. For example, \cite{8792076_endtoend} provides an iterative algorithm for the training of communication systems with an unknown channel model or with non-differentiable components.

Potential future research directions include its practical implementation aspects in terms of when and where to train the models, how to accompany varying channel statistics in long-term, transmitter-receiver synchronization, etc.

\subsubsection{Joint Channel Estimation and Detection}

Channel estimation, equalization, and signal detection are three tasks in the physical layer that are relatively easy to carry out individually. However, the optimal joint design is substantially more complicated, making ML a potential shortcut.
In \cite{8052521_channel_est}, a learning-based joint channel estimation and signal detection algorithm is proposed for OFDM systems, which can reduce the signaling overhead and deal with nonlinear clipping. In \cite{8933050}, joint channel estimation and signal detection is carried out when both the transmitter and receiver are subject to hardware impairments with an unknown model.
Online learning-based channel estimation and equalization is proposed in \cite{8715649_channel_est}, which can jointly handle fading channels and nonlinear distortion.

Methods to overcome the challenge of offline model training, which causes performance degradation due to the discrepancies between the real channels and training channels, need to be considered. ML implementations with online training and constructing training data to match real-world channel conditions are potential future directions in this regard.

\subsubsection{Resource Management}

Many resource management problems are combinatorial in nature, which implies that optimal exhaustive-search based algorithms are impossible to utilize in practice. This gives an opportunity for ML-based algorithms to outperform existing suboptimal approaches.
A generic resource allocation framework is provided in \cite{8680025}. A specific example is hybrid precoder-combiner design in multi-user mmWave systems \cite{8890805_hybrid_bf}, where the hardware is incapable of the optimal digital processing.

Some potential research areas which are already being studied include power control, beamforming in massive MIMO and in cell-free environments, predictive scheduling and resource allocation, etc. Reinforcement learning frameworks and transfer learning techniques to learn, adapt, and optimize for varying conditions over time are expected to be useful in performing resource management tasks with minimal supervision \cite{6GFlagship_ML_WP}.

%-------------------------------------------------------------------------
\subsection{Coded Caching}

Coded caching (CC) was proposed in \cite{maddah2014fundamental} as a way to increase the data rate, with the help of cache memories available throughout the network. It enables a global caching gain, proportional to the total cache size of all network users, to be achieved in addition to the local caching gain at each user. This additional gain is achieved by multicasting carefully created codewords to various user groups, such that each codeword contains useful data for every user in the target group. Technically, if $K$, $M$ and $N$ denote the user count, cache size at each user and the file library size respectively, using CC the total data size transmitted over the broadcast link can be reduced by a factor of $1+t$, where $t=\frac{KM}{N}$ is called the CC gain. 

Interestingly, the CC gain is not only achievable in multi-antenna communications but is also additive with the spatial multiplexing gain of using multiple antennas~\cite{shariatpanahi2018physical}.
This nice property is achieved through multicast beamforming of multiple parallel (partially overlapping) CC codewords to larger sets of users, while removing (or suppressing) inter-codeword interference with the help of carefully designed beamforming vectors \cite{tolli2017multi}.

\begin{figure}[t]
\begin{center}
    \includegraphics[scale=0.55]{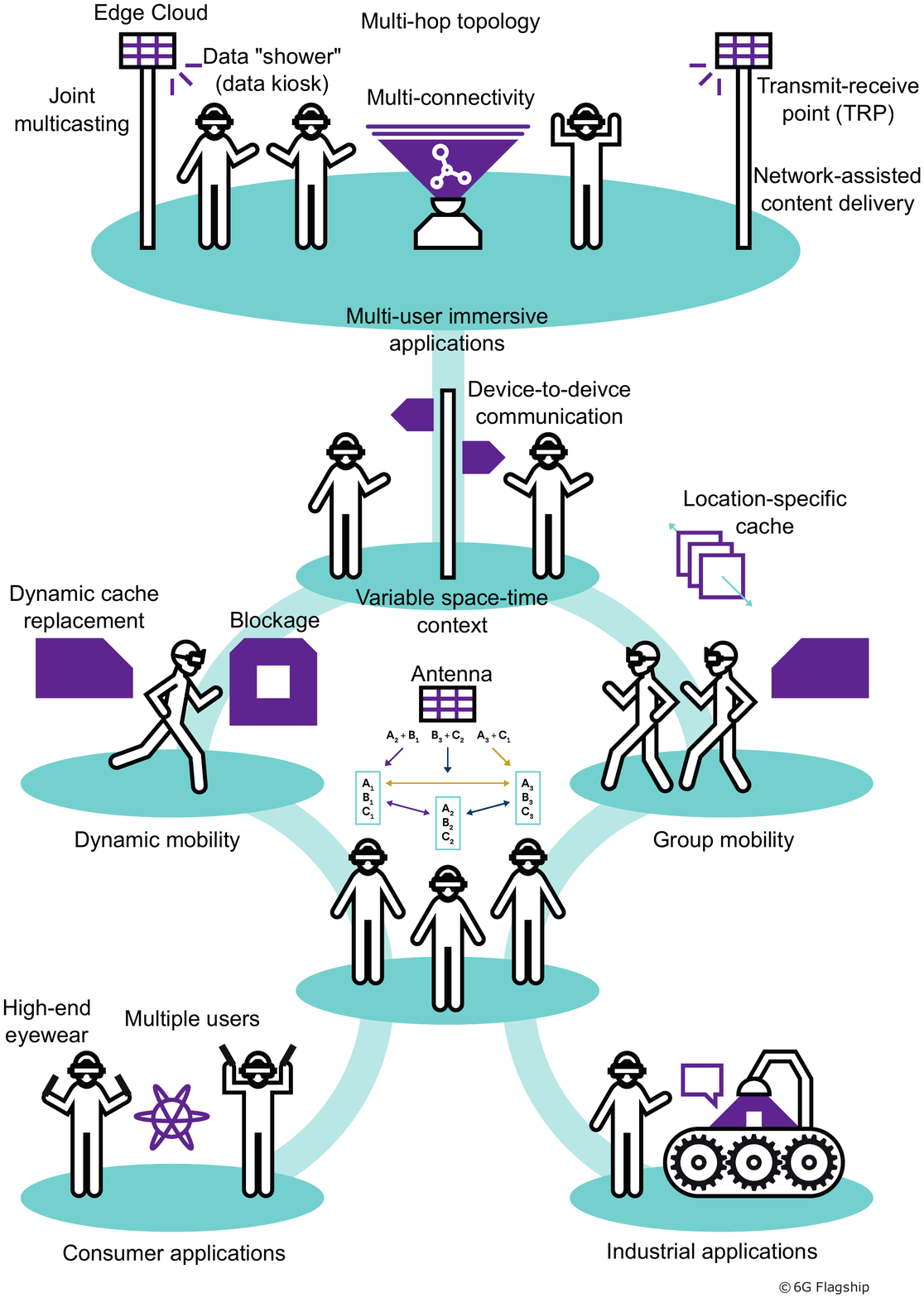}
    \caption{Immersive viewing scenario with coded caching.}
    \label{fig:immersive_vieweing}
\end{center}
\end{figure}

The number of multimedia applications benefiting from CC is expected to grow in the future. One envisioned scenario assumes an extended reality or hyper-reality environment (e.g., educational, industrial, gaming, defense, social networking), as depicted in Figure~\ref{fig:immersive_vieweing}. A large group of users is submerged into a network-based immersive application, which runs on a high-end eye-wear, that requires heavy multimedia traffic and is bound to guarantee a well-defined quality-of-experience level for every user in the operating theatre. 
In such a scenario, a possibility of caching at the user devices and computation offloading onto the network edge opens possibilities to deliver high-throughput low-latency traffic, while ensuring its stability and reliability for a truly immersive experience. The fact that modern mobile devices continue to grow their storage capacity (it is one of the cheapest network resources) makes CC especially beneficial given the uniqueness of this use case, where the popularity of limited and location-dependent content becomes much higher than in any traditional network. 

Alongside its strong features, there are many practical issues with CC that must be addressed to enable implementation in 6G. Due to its structure, CC is heavily dependent on the underlying multicasting implementation. Designing efficient multicast and broadcast strategies for 6G is part of the ongoing research (cf.~Section \ref{section:broadcast_multicast}).
On the other hand, the large subpacketization requirement~\cite{lampiris2018adding, salehi2019codedlinear}, poor performance in the presence of users with diverse channel conditions, performance dependency on the complexity of the beamforming structure~\cite{salehi2019beamformer} and the resource allocation and network control overhead are important open issues which should be solved for a practical CC implementation. 
All these issues are now under active research by the community.

%-------------------------------------------------------------------------
\subsection{Full-Coverage Broadband Connectivity} 

Around four billion people largely lack broadband connectivity. In order to reduce the global digital divide, the 6G research community needs to vigorously address the largely neglected problem of full-coverage broadband access for people and devices in rural regions and at remote locations. However, each new wireless generation has unfortunately increased the digital divide between urban and rural areas, rather than reduced it. While new advanced access technologies have given huge benefits, especially in densely populated regions, they have offered much fewer benefits in rural regions and possibilities to connect to remote locations. Furthermore, new wireless generations have often resulted in reduced area coverage than previous generations.

Recent and emerging wireless technologies at millimeter and THz frequencies are primarily targeting to improve the wireless connectivity where it is already relatively good; that is, at short distances to the access network and to serve densely packed users and devices with even more capacity. The rush towards higher peak data rates has the by-product of increasing the digital divide by offering most gain to the users and devices which are physically near to an abundant radio access infrastructure. 

One potential exception is massive MIMO, which primarily improves the performance at the cell edge. While conventional high-gain sector antennas provide array gains in the cell center, the highly adaptive beamforming of massive MIMO can provide arrays gains wherever the user is located in the coverage area. Massive MIMO is already being used to provide fixed wireless access in suburban areas, and \cite[Sec.~6.1]{Marzetta2016a} demonstrates how it can evolve providing 3000 homes with broadband access in a cell with an 11 km radius. However, such theoretical solutions might only become a reality if they become economically viable. Due to the unavoidable pathloss difference in a cell, the user fairness balancing, given limited resources, comes at the price of substantially lower sum-rate and provides considerably less capacity for the oftentimes many more short-distance users that pay for the network. It is uncertain if the network operators are willing to utilize the current massive MIMO technology in rural areas and to configure it to achieve user fairness.

For such reasons, the research community needs to focus also on developing access methods and broadband connectivity systems that inherently offer as uniform capacity as possible over very large areas, while the cost per user is as low as possible. However, this is obviously a huge challenge in practice due to the oftentimes poor channel conditions in rural areas and remote locations.  
  
Three potential ways of utilizing the aforementioned technologies to bridge the digital divide are described next.

\begin{itemize}
\item \textbf{Long-range massive MIMO.} \emph{Potential:} Can greatly increase the range and coverage of an AP and increase the capacity also at long distances, e.g., using very tall towers. Coherent joint transmission between adjacent APs can be utilized to enhance the performance and diversity at cell edges, effectively creating a long-range cell-free massive MIMO system.
\emph{Challenges:} Long-distance LOS dominated channels are more sensitive to shadowing and blocking, and frequent slow fading in general, which impose practical challenges for massive MIMO in rural areas. Unfortunate homes and places may be in a permanent deep shadow and thus always out of coverage. The near-far problem exists and often enhanced due to the nature of the wireless channels, with typically decreasing spatial diversity at longer distances. Coherent joint transmission from multiple APs requires more backhaul/fronthaul capacity and is subject to large delays. The synchronization of widely distributed transmitters is also very challenging due to large variations in propagation delays.

\item \textbf{mmWave and THz frequencies from satellites.} \emph{Potential:} LEO-satellite constellations offer larger coverage and capacity in rural and remote areas and can be used in conjunction with high-altitude platform stations (HAPS). Since these solutions view the coverage area from above, they can fill in the coverage gaps of long-range massive MIMO, as described above, and also offer large-capacity backhaul links for ground-based APs. High array gains are possible by using many antenna elements on the satellites and HAPS. Each user can either be served by the best available satellite/HAPS, or by coherent transmission from multiple ones. The coordination can be enabled via optical inter-satellite connections. Coherent transmission from satellites and ground-based APs is also possible, to effectively create a space-based cell-free network.
\emph{Challenges:} A large number of LEO-satellites are required to make them constantly available in all rural regions, which is associated with a high deployment cost. Interference may arise between uncoordinated satellites and terrestrial systems using the same radio spectrum. The large pathloss per antenna element requires large antenna arrays and adaptive beamforming, since satellites are constantly in motion. The transmit power is limited, particularly in the uplink. Coherent transmission from multiple satellites, or between space and ground, requires higher fronthaul capacity and synchronization and causes delays.

\item \textbf{Intelligent reflecting surfaces.} \emph{Potential:} Can be deployed on hills and mountains to remove coverage holes in existing networks by reflecting signals from APs towards places where the natural multipath propagation is insufficient. The IRS can be powered by solar panels or other renewable sources. The cost is low compared to deploying and operating additional APs since no backhaul infrastructure or connection to the power grid is needed.
\emph{Challenges:} The larger the propagation distance, the larger the surface area needs to be in order to become effective. Remote control of the IRS is challenging and might only support fixed access or low mobility. An IRS deployed in nature is more prone to sabotage than hardware deployed in tall masts.

\end{itemize}

%=========================================================================
\section{Summary: New Concepts for 6G from a Broadband Connectivity Point of View}
%=========================================================================

Based on the topics elaborated earlier, the following novel directions are highlighted and summarized in Table~\ref{tab:summary}. 

\begin{table}[h]
\centering
\begin{tabular}{ |p{5cm}|p{5cm}|p{5cm}|  }
\hline
\multicolumn{3}{|c|}{\textbf{Summary of key open problems}} \\
\hline
\textbf{Challenges} & \textbf{Potential 6G solutions} & \textbf{Open research questions} \\
\hline
Stable service quality in coverage area & User-centric cell-free massive MIMO  & Scalable synchronization, control, and resource allocation \\
\hline
Coverage improvements & Integration of a spaceborne layer, ultra-massive MIMO from tall towers, intelligent reflecting surfaces & Joint control of space and ground based APs, real-time control of IRS \\
\hline
Extremely wide bandwidths & Sub-THz, VLC & Hardware development and mitigation of impairments \\
\hline
Reduced latency    & Faster forward error correcting schemes, wider bandwidths & Efficient encoding and decoding algorithms \\
\hline
Efficient spectrum utilization & Ultra-massive MIMO, waveform adaptation, interference cancellation & Holographic radio, use-case based waveforms, full-duplex, rate-splitting\\
\hline
Efficient backhaul infrastructure & Integrated access and backhauling & Dynamic resource allocation framework using space and frequency domains \\
\hline
Smart radio environment &  Intelligent reflecting surfaces & Channel estimation, hardware development, remote control \\
\hline
Energy efficiency & Cell-free massive MIMO, suitable modulation techniques  & Novel modulation methods with limited hardware complexity  \\
\hline
Modeling or algorithmic deficiencies in complex and dynamic scenarios & ML/AI based model-free, data-driven learning and optimization techniques & End-to-end learning/joint optimization, unsupervised learning for radio resource management \\
%Integrated connectivity &  &  \\
\hline
\end{tabular}
\caption{Summary of key open problems.}
\label{tab:summary}
\end{table}

There will be a paradigm change in the way users are supported, shifting from the network-centric view where networks are deployed to deliver extreme peak rates in special cases to a user-centric view where consistently high rates are prioritized. Such ubiquitous connectivity can be delivered by  cell-free massive MIMO, IAB, and complemented by IRSs. While the sub-6 GHz spectrum will define the wide-area coverage and spatial multiplexing, a significant effort is expected in sub-THz and THz bands to achieve short-range connectivity with data rates in the vicinity of 1~Tbps. Novel coding, modulation and waveforms will be needed to support extreme data rates with manageable complexity and robustness to the hardware impairments that increase with the carrier frequency. In situations where the beamforming capabilities offered by the ultra-massive MIMO technology are insufficient to manage interference in the spatial domain, coding methods based on rate-splitting or broadcasting can be also utilized. The efficiency can be also improved by making use of caching techniques. Moving on to photonics-based methods, one key development can be in holographic radio. This will revolutionize the way the communication is carried out until now. Seamless integration between satellites and terrestrial networks will ensure that large populations outside the urban centers will be reached by high-quality broadband connectivity.

% \newpage

%=========================================================================
\section*{Acknowledgements}
%=========================================================================

This white paper has been written by an international group of experts led by the Finnish 6G~Flagship program at the University of Oulu (https://www.oulu.fi/6gflagship). It is part of a series of twelve 6G white papers to be published in their final format in June 2020.

%=========================================================================
\footnotesize
\bibliographystyle{IEEEtran}
\bibliography{IEEEabrv,bibliography}
%=========================================================================

\end{document}